\documentclass[vecphys]{svmult}

\usepackage{makeidx}         % allows index generation
\usepackage{graphicx}        % standard LaTeX graphics tool
\usepackage{multicol}        % used for the two-column index
\usepackage[sort]{cite}            % adjusts the "syntax" of the refs in the text
\usepackage[bottom]{footmisc}% places footnotes at page bottom
\usepackage{amsfonts,amsmath,amssymb}
\usepackage[english]{babel}

\usepackage{xspace}
\usepackage{xcolor}
\usepackage[latin1]{inputenc}
\usepackage{placeins}
\usepackage{floatflt}
\usepackage[caption=false]{subfig}

\newcommand{\Pb}{\mbox{\rm (P)}\xspace}

\newcommand{\Pbn}{\mbox{\rm (P$_\nu$)}\xspace}
\newcommand{\Pbsp}{\mbox{\rm (P$_{\mbox{sp}}$)}\xspace}

\newcommand{\Uad}{\mathcal{F}_{{\rm ad}}}

\newcommand{\R}{\mathbb{R}}

\newcommand{\eq}[1]{$\mathrm{Eq.}$~\eqref{#1}}
\newcommand{\eqs}[1]{$\mathrm{Eqs.}$~\eqref{#1}}
\newcommand{\bracket}[1]{\left(#1\right)}

\newcommand{\ltwoq}{{L^2(Q)}}
\newcommand{\ltwo}{{L^2(\Omega)}}
\newcommand{\intQb}[1]{\int\limits_0^T \hspace*{-5pt} \int\limits_\Omega d\vec{x}\,dt\,\left\{ #1 \right\}}
\newcommand{\intOmb}[1]{\int\limits_\Omega d\vec{x}\,\left\{ #1 \right\}}

\makeindex             % used for the subject index
\begin{document}

\title*{Analytical, Optimal, and Sparse Optimal Control of Traveling Wave Solutions to Reaction-Diffusion Systems}
\titlerunning{Analytical and Optimal Position Control of Reaction-Diffusion patterns}
\author{Christopher Ryll\inst{1} \and Jakob L\"ober\inst{2} \and Steffen Martens\inst{2} \and Harald Engel\inst{2} \and Fredi Tr\"oltzsch\inst{1}}
\authorrunning{ Christopher Ryll et al. }
\institute{Technische Universit\"at Berlin, Institut f\"ur Mathematik, 10623 Berlin, Germany
      \and Technische Universit\"at Berlin, Institut f\"ur Theoretische Physik, 10623 Berlin, Germany}
\maketitle

\begin{abstract}
This work deals with the position control of selected patterns in
reaction-diffusion systems. Exemplarily, the Schl\"{o}gl and
FitzHugh-Nagumo model are discussed using three different approaches.
First, an analytical solution is proposed. Second, the standard optimal
control procedure is applied. The third approach extends
standard optimal control to so-called sparse optimal control that results in
very localized control signals and allows the analysis of second order
optimality conditions.
\end{abstract}
 
\section*{Introduction}
\noindent Beside the well-known Turing patterns, reaction-diffusion (RD) systems possess a rich variety of self-organized spatio-temporal wave patterns including propagating fronts, solitary excitation pulses, and periodic pulse trains in one-dimensional media. These patterns are ``building blocks'' of wave patterns like target patterns, wave segments, and spiral waves in two as well as scroll waves in three spatial dimensions, respectively. Another important class of RD patterns are stationary, breathing, and moving localized spots \cite{schloegl72,Winfree1972,tyson1988singular,kapral1995chemical,Kuramoto2003,Murray2003,liehr2013dissipative}.

Several control strategies have been developed for purposeful manipulation of wave dynamics as the application of closed-loop or feedback-mediated control loops with and without
delays \cite{Zykov2004,ZykovPhysD2004,Mikhailov2006,Schlesner2006} and open-loop control that includes external spatio-temporal forcing \cite{Kim2001,Mikhailov2006,Zykov2003,Chen2009}, optimal control \cite{elma86,bjt10,buch_eng_kamm_tro2013}, and control by imposed geometric constraints and heterogeneities on the medium \cite{Haas1995,MartensPRE2015}.
While feedback-mediated control relies on continuously monitoring of the system's state, open-loop control is based on a detailed knowledge of the system's dynamics and its parameters.

Experimentally, feedback control loop have been developed for the photosensitive Belousov-Zhabotinsky (BZ) reaction. 
%\cite{Zhabotinsky1973,winfree1973scroll} 
The feedback signals are obtained from wave activity measured at one or several detector points, along 
detector lines, or in a spatially extended control domain including global feedback control 
\cite{Zykov2004,ZykovPhysD2004,Zykov2005}. Varying the excitability of the light-sensitive BZ medium by 
changing the globally applied light intensity
%, being an example for spatio-temporal forcing,
forces a spiral wave tip to describe a wide range of hypocycloidal and epicycloidal trajectories 
\cite{steinbock1993control,Schrader1995}. Moreover, feedback-mediated control loops have been applied 
successfully in order to stabilize unstable patterns in experiments, such as unstable traveling wave segments
and spots \cite{Schlesner2006}. Two feedback loops were used to guide unstable wave segments in the BZ reaction along pre-given trajectories \cite{sakurai2002}. An open loop control was successfully
deployed in dragging traveling chemical pulses of adsorbed CO during heterogeneous catalysis on platinum single crystal surfaces \cite{WolffPRL20003}. In these experiments, the
pulse velocity was controlled by a laser beam creating a movable localized temperature heterogeneity on an addressable catalyst surface, resulting in a V-shaped pattern
\cite{wolffScience2001}. Dragging a one-dimensional chemical front or phase interface to a new position by anchoring it to a movable parameter heterogeneity, was studied theoretically
in \cite{malomed2002pulled,Kevrekidis2004}.

Recently, an open-loop control for controlling the position of traveling waves over time according to 
a prescribed \emph{protocol of motion} $\vec{\phi}(t)$ was proposed that preserves simultaneously 
the wave profile \cite{loeber_engel2014}. Although position control is realized by external spatio-temporal 
forcing, i.e., it is an open-loop control, no detailed knowledge about the reaction dynamics 
as well as the system parameters is needed.  We have already demonstrated the ability of position control to 
accelerate or decelerate traveling fronts and pulses in one spatial dimension for a variety of RD models 
\cite{lober2014stability,LoeberBook2014}. In particular, we found  that the analytically derived control 
function is close to a numerically obtained optimal control solution. A similar approach allows to control 
the two-dimensional shape of traveling wave solutions. Control signals that realize a desired wave shape are 
determined analytically from nonlinear evolution equations for isoconcentration lines as the perturbed nonlinear 
phase diffusion equation or the perturbed linear eikonal equation \cite{lober2014shaping}.
%
%Additionally, we extended the approach to 
%two spatial dimensions which allows to dynamically shape wave patterns by prescribing a %time-dependent contour \cite{lober2014shaping}. 
% 
In the work at hand, we compare our analytic approach for position control with optimal trajectory tracking of 
RD patterns in more detail. In particular, we quantify the difference between an analytical solution and a 
numerically obtained result to optimal control. Thereby, we determine the conditions under which the numerical 
result approaches the analytical result. This establishes a basis for using analytical solutions to speed up numerical computations of optimal control and serves as a consistency check for numerical algorithms.
\medskip

\index{reaction-diffusion systems}
\noindent We consider the following controlled RD system
\begin{equation}
 \partial_t \vec{u}(\vec{x},t) -  \vec{\mathcal{D}} \Delta \vec{u}(\vec{x},t) + \vec{R}(\vec{u}(\vec{x},t)) 
  = \vec{\mathcal{B}}\vec{f}(\vec{x},t)\label{eq:RDS}.
\end{equation}
Here, $\vec{u}(\vec{x},t)=(u_1(\vec{x},t),\dots,u_n(\vec{x},t))^T$ is a vector of $n$ state components in a bounded or unbounded spatial
domain $\Omega \subset \R^N$ of dimension $N \in \{1,2,3\}$. $\vec{\mathcal{D}}$ is an $n\times n$ matrix of 
diffusion coefficients which is assumed to be diagonal, $\vec{\mathcal{D}}=\mathrm{diag}(D_1,\ldots,D_n)$, because the medium is presumed to be isotropic. 
$\Delta$ represents the $N$-dimensional Laplacian operator, and 
$\vec{R}$ denotes the vector of $n$ reaction kinetics which, in general, are nonlinear functions of the state. The vector of control signals 
$\vec{f}(\vec{x},t)=(f_1(\vec{x},t),\dots,f_m(\vec{x},t))^T$ acts at all times and everywhere within the spatial domain $\Omega$. The 
latter assumption is rarely justified in experiments, where the application of control signals is often restricted to subsets of $\Omega$.
However, notable exceptions, as e.g. the already mentioned photosensitive BZ reaction, exist. Here, the light intensity is deployed as 
the control signal such that the control acts everywhere within a two-dimensional domain.

Equation \eqref{eq:RDS} must be supplemented with an initial condition $\vec{u}(\vec{x},t_0)=\vec{u}_0(\vec{x})$ and appropriate 
boundary conditions. A common choice are no-flux boundary 
conditions at the boundary $\Sigma = \partial\Omega \times (0,T)$, $\partial_n \vec{u}(\vec{x},t) = \vec{0}$, 
where $\partial_n \vec{u}$ denotes the component-wise
spatial derivative in the direction normal to the boundary $\Gamma = \partial\Omega$ of the spatial domain.

Typically, the number $m$ of independent control signals in \eq{eq:RDS} is smaller than the number $n$ of state components. We call such a 
system an \emph{underactuated} system. 
The $n\times m$ matrix $\vec{\mathcal{B}}$ determines which state components are directly affected by the control signals. If $m=n$ 
and the matrix $\vec{\mathcal{B}}$ is regular, it is called a \emph{fully actuated} system.
% \medskip

Our main goal is to identify a control $\vec{f}$ such that the state $\vec{u}$ follows a desired spatio-temporal trajectory 
$\vec{u}_d$, also called a desired distribution, as closely as possible everywhere in space $ \Omega $ and for all times 
$0 \leq t\leq T$. We can measure the distance between the actual solution 
$\vec{u}$ of the controlled RD system \eq{eq:RDS} and the desired trajectory $\vec{u}_d$ up to the 
terminal time $T$ with the non-negative functional
\begin{align}
% J(\vec{u}) = \intQb{(\vec{u}(\vec{x},t) - \vec{u}_d(\vec{x},t))^2},
J(\vec{u}) =& \|\vec{u} - \vec{u}_d\|_\ltwoq^2, \label{eq:Distdist}
\intertext{where $\| \cdot \|_\ltwoq^2$ is the $L^2$-norm defined by}
\| \vec{h} \|_\ltwoq^2 =& \intQb{\vec{h}(\vec{x},t)^2 }, \label{eq:L2norm}
\end{align}
in the space-time-cylinder $Q := \Omega \times (0,T)$. The functional \eq{eq:Distdist} reaches its smallest possible value, $J=0$, 
if and only if the controlled state $\vec{u}$ equals 
the desired trajectory almost everywhere in time and space.

In many cases, the desired trajectory $\vec{u}_d$ cannot be realized exactly by the control, cf. Ref. \cite{lober2015thesis} for examples. However, one might be able to find a control which enforces the state $\vec{u}$ to follow
$\vec{u}_d$ as closely as possible as measured by $J$. A control $\vec{f} = \vec{\bar f}$ is \emph{optimal} if it realizes a 
state $\vec{u}$ which minimizes $J$.
The method of optimal control views $J$ as a constrained functional subject to $\vec{u}$ satisfying the 
controlled RD system \eq{eq:RDS}.

Often, the minimum of the objective functional $J$, \eq{eq:Distdist}, does not exist within appropriate function spaces. 
Consider, for example, the assumption that the controlled state, obtained as a solution to the optimization problem, is continuous in time and space. Despite that a discontinuous state $\vec{u}$ leading to a smaller value for $J(\vec{u})$ than any continuous function might exist, this state is not regarded as a solution to the optimization problem. Furthermore, a control enforcing a discontinuous state may diverge at exactly the points of discontinuity; examples in the context of dynamical systems are discussed in Ref. \cite{lober2015thesis}. For that reason, the \emph{unregularized} optimization problem, \eq{eq:Distdist}, is also called a singular optimal control problem.To ensure the existence of a minimum of $J$ and bounded control signals, additional (inequality) constraints such as bounds for the control signal can be introduced, cf. Ref. \cite{hlt98}. Alternatively, it is possible to add a so-called Tikhonov-regularization term to the functional \eq{eq:Distdist} which is quadratic in the control,
\begin{equation}
% J(\vec{u},\vec{f}) = \intQb{(\vec{u}(\vec{x},t) - \vec{u}_d(\vec{x},t))^2 + \nu \vec{f}^2(\vec{x},t)}.
 J(\vec{u},\vec{f}) = \|\vec{u} - \vec{u}_d\|_\ltwoq^2 + \nu \|\vec{f}\|_\ltwoq^2.
 \label{eq:DistdistRegularized}
\end{equation}
The $L^2$-norm of the control $\vec{f}$ is weighted by a small coefficient $\nu>0$. This term might be interpreted as a 
control cost to achieve a certain state $\vec{u}$. Since the control $\vec{f}$ does not come for free, there is a ``price'' to pay. 
In numerical computations, $\nu>0$ serves as a regularization parameter that stabilizes the algorithm. For the numerical results 
shown in later sections, we typically choose $\nu$ in the range $10^{-8} \leq \nu \leq 10^{-3}$. While $\nu>0$ guarantees the 
existence of an optimal control $\vec{f}$ in one and two spatial dimensions even in the absence of bounds on the control signal \cite{hlt98}, it is not known whether Tikhonov-regularization alone also works in spatial dimensions $N$ larger than two. Here,
we restrict our investigations to one and two spatial dimensions. The presence of the regularization term causes the states to be further 
away from the desired trajectories than in the case of $\nu = 0$. Thus, the case $\nu=0$ is of special interest. Naturally, 
the solution $\vec{u}$ for $\nu = 0$ is the closest (in the $L^2(Q)$-sense) to the desired trajectory $\vec{u}_d$ among all 
optimal solutions associated with any $\nu \geq 0$. Therefore, it can be seen as the limit of realizability of a certain desired 
trajectory $\vec{u}_d$.

% Usually, this coefficient attains values in between $10^{-8}\leq\nu\leq10^{-3}$. In some applications, this additional term can be understood as 
% the cost to achieve a certain state $\vec{u}$.
% Although a finite value $\nu>0$ is often necessary for the analysis of the problem on the one hand and increases the stability 
% of numerical computations of an optimal control on the other 
% hand, the case $\nu=0$ is of special interest. Naturally,
% the solution $\vec{u}$ for $\nu = 0$ - if it exists - is the closest (in the $L^2(Q)$-sense) to the desired trajectory 
% $\vec{u}_d$ among all optimal solutions associated with any $\nu \geq 0$. Therefore, it can be seen as the limit of the 
% realizability of a certain desired trajectory $\vec{u}_d$.

In addition to the weighted $L^2$-norm of the control, other terms can be added to the functional \eq{eq:DistdistRegularized}.
An interesting choice is the weighted $L^1$-norm such that the functional reads
\begin{equation}
 %J(\vec{u},\vec{f}) = \intQb{(\vec{u}(\vec{x},t) - \vec{u}_d(\vec{x},t))^2 + \nu \vec{f}^2(\vec{x},t) + \kappa |\vec{f}(\vec{x},t)|},
J(\vec{u},\vec{f}) = \|\vec{u} - \vec{u}_d\|_\ltwoq^2 + \nu\|\vec{f}\|_\ltwoq^2 + \kappa \|\vec{f}\|_{L^1(Q)}.
 \label{eq:DistdistSparse}
\end{equation}
For appropriate values of $\kappa>0$, the corresponding optimal control becomes sparse, i.e., it only acts in some 
localized regions of the space-time-cylinder while it vanishes identically everywhere else. 
Therefore, it is also called \emph{sparse control} or \emph{sparse optimal control} in the literature, see Refs. 
\cite{stadler2009,wachsmuth_wachsmuth2011,CHW2012a,casas_troeltzsch_sparse_state2013}.
In some sense, we can interpret the areas with non-vanishing sparse optimal control signals as the most sensitive areas of the 
RD patterns with respect to the desired control goal. A manipulation of the RD pattern in these areas is most efficient while
control signals applied in other regions have only weak impact. Furthermore, the weighted $L^1$-norm enables the 
analysis of solutions with a Tikhonov-regularization parameter $\nu$ tending to zero. This allows to draw 
conclusions about the approximation of solutions to unregularized problems by regularized ones.

\medskip

In Sect. \ref{S1}, we present an analytical approach for the control of the position of RD patterns in fully actuated systems.
These analytical expressions are solutions to the unregularized ($\nu=0$) optimization problem, \eq{eq:Distdist}, and might provide 
an appropriate initial guess for numerical optimal control algorithms. Notably, neither the controlled state nor the control signal 
suffering from the problems are usually associated with unregularized optimal control; both expressions yield continuous and bounded solutions under certain assumptions postulated in Sect. \ref{S1}. 
In Sect. \ref{S2}, we state explicitly the optimal control problem for traveling wave solutions to the Schl\"ogl \cite{schloegl72,Zeldovich1938} and 
the FitzHugh-Nagumo model \cite{nagumo62,fitz61}. Both are well-known models to describe traveling fronts and pulses 
in one spatial dimension, solitary spots and spiral waves in two spatial dimensions, and scroll waves in three spatial dimensions 
\cite{kapral1995chemical,Mikhailov2006,AzhandEPL2014,Totz2015}. We compare the analytical solutions from Sect. \ref{S1} with a numerically obtained regularized optimal control solution for the position control of a traveling front solution in the one-dimensional Schlögl model in Sect. \ref{S2-3}. 
In particular, we demonstrate the convergence of the numerical result to the analytical solution 
for decreasing values $\nu$. The agreement becomes perfect within numerical accuracy if $\nu$ is chosen sufficiently small. Section \ref{S3} discusses sparse optimal 
control in detail and presents numerical examples obtained for the FitzHugh-Nagumo system. Finally, we conclude our findings in Sect. \ref{S4}.

\section{Analytical approach}\label{S1}
\index{position control}
Below, we sketch the idea of analytical position control of RD patterns proposed previously in Refs. 
\cite{loeber_engel2014,lober2014stability}. For simplicity, we consider a single-component 
RD system of the form
\begin{equation}
\displaystyle\ \partial_t u(x,t) - D \partial_x^2 u(x,t) + R(u(x,t)) = f(x,t),\label{eq:SchloeglModel} 
\end{equation}
in a one-dimensional infinitely extended spatial domain $x \in \R$. The state $u$ as well as the control 
signal $f$ are scalar functions and the system \eq{eq:SchloeglModel} is fully actuated. Usually, \eq{eq:SchloeglModel}
is viewed as a differential equation for the state $u$ with the control signal $f$ acting as an inhomogeneity. Alternatively,
\eq{eq:SchloeglModel} can also be seen as an expression for the control signal. Exploiting this
relation, one simply inserts the desired trajectory $u_d$ for $u$ in \eq{eq:SchloeglModel} and obtains for the control
\begin{equation}
 f(x,t)=\displaystyle \partial_t u_d(x,t) - D \partial_x^2 u_d(x,t) + R(u_d(x,t)).\label{eq:ControlSolution}
\end{equation}
In the following, we assume that the desired trajectory $u_d$ is sufficiently smooth everywhere in the space-time-cylinder $Q$ such that the evaluation 
of the derivatives $\partial_t u_d$ and $\partial_x^2 u_d$ yields continuous expressions.
We call a desired trajectory $u_d$ \emph{exactly realizable} if the controlled state $u$ equals $u_d$ everywhere in $Q$, i.e., $u(x,t) = u_d(x,t)$.
For the control signal given by \eq{eq:ControlSolution}, this can only be true if two more conditions are satisfied. First, the initial 
condition for the controlled state must coincide with the initial state of the desired trajectory, i.e., $u(x,t_0) = u_d(x,t_0)$. 
Second, all boundary conditions obeyed by $u$ have to be obeyed by the desired trajectory $u_d$ as well.
Because  of $u(x,t) = u_d(x,t)$, the corresponding unregularized functional $J$, \eq{eq:Distdist}, vanishes identically. Thus, the control 
$f$ is certainly a control which minimizes the unregularized functional $J$ and, in particular, it is optimal.

In conclusion, we found a solution to the unregularized optimization problem \eq{eq:Distdist}. The solution for the  controlled state is 
simply $u(x,t) = u_d(x,t)$, while the solution for the control signal is given by \eq{eq:ControlSolution}. Even though we are 
dealing with an unregularized optimization problem, the control signal as well as the controlled state are continuous and bounded 
functions provided the desired trajectory $u_d$ is sufficiently smooth in space and time.
%A more rigorous evaluation 
%based on the necessary optimality condition for the minimization of the regularized functional, \eq{eq:DistdistRegularized}, 
%leads to the same conclusion \cite{lober2015thesis}.
% If $\nu=0$ and $u(x,t) = u_d(x,t)$ everywhere in space and time, including both the boundary of the domain and the initial state, 
% the control solution \eq{eq:ControlSolution} satisfies all necessary optimality conditions \cite{lober2015thesis}.

Generalizing the procedure to multi-component RD systems in multiple spatial dimensions, the expression for the control reads
\begin{equation}
 \vec{f}(x,t) = \vec{\mathcal{B}}^{-1}(\displaystyle \partial_t \vec{u}_d(x,t) 
  - \vec{\mathcal{D}} \Delta \vec{u}_d(x,t) + \vec{R}(\vec{u}_d(x,t))).\label{eq:ControlSolutionMultiComponent}
\end{equation}
Once more, the initial and boundary conditions for the desired trajectory $\vec{u}_d$ have to comply with the initial and boundary conditions 
of the state $\vec{u}$.
Clearly, the inverse of $\vec{\mathcal{B}}$ exists if and only if $\vec{\mathcal{B}}$ is a regular square matrix, i.e., the system 
must be fully actuated. We emphasize the generality of the result. Apart from mild conditions
on the smoothness of the desired distributions $\vec{u}_d$, \eq{eq:ControlSolutionMultiComponent} yields a simple expression for the control signal for arbitrary
$\vec{u}_d$.
% In other words, the number of independent control signals must equal the number of state components, i.e., the system must be fully actuated.

Next, we consider exemplarily the position control of traveling waves (TW) in one spatial dimension. 
Traveling waves are solutions to the uncontrolled RD system, i.e., \eq{eq:RDS} with $\vec{f}= \vec{0}$.
They are characterized by a wave profile $\vec{u}(x,t)=\vec{U}_c(x-c\,t)$ which is stationary in a 
frame of reference $\xi=x-ct$ co-moving with velocity $c$. The wave profile $\vec{U}_c$
satisfies the following ordinary differential equation (ODE),
\begin{equation}
 \displaystyle \vec{\mathcal{D}} \vec{U}_c''(\xi)+c\vec{U}_c'(\xi) - \vec{R}(\vec{U}_c(\xi)) 
 = \vec{0},\quad \xi \in \Omega \subset \R. \label{eq:ProfileEquation}
\end{equation}
The prime denotes differentiation with respect to $\xi$. Note that stationary 
solutions with a vanishing propagation velocity  $c = 0$ are also considered as traveling waves. 
The ODE for the wave profile, Eq.~\eqref{eq:ProfileEquation}, can exhibit one or more homogeneous steady states. 
Typically, the wave profile $\vec{U}_c$ approaches either two different steady states or the same steady state for $\xi \to \pm \infty$. 
This fact can be used to classify traveling wave profiles. Front profiles connect different steady states for $\xi \to \pm \infty$ 
and are found to be heteroclinic orbits of Eq.~\eqref{eq:ProfileEquation}. Pulse profiles join the same steady state and are 
found to be homoclinic orbits \cite{guckenheimer1983nonlinear}.
% Typically, the wave profile $\vec{U}_c$ approaches either two different steady states (front profiles) or the same steady state 
% (pulse profiles) for $\xi \to \pm \infty$. 
Furthermore, all TW solutions are localized in the sense that their spatial derivatives of any order $m\geq 1$ decay to zero, 
$\lim_{\xi \to \pm \infty} \partial_\xi^m \vec{U}_c(\xi)=0$. 
% Thus, TW solutions only exists in infinite domains $\Omega = \R$.

We propose a spatio-temporal control signal $\vec f(x,t)$ which shifts the traveling wave according to a prescribed 
protocol of motion $\phi(t)$ while simultaneously preserving the uncontrolled wave profile $\vec{U}_c$. 
Correspondingly, the desired trajectory reads 
\begin{equation}
\vec{u}_d(x,t)=\vec{U}_c(x-\phi(t)). \label{eq:PositionControlDesiredTrajectory}
\end{equation}
Note that the desired trajectory is localized for all values of $\phi(t)$ because the TW profile $\vec{U}_c$ is localized.
The initial condition for the state is $\vec{u}(x,t_0)=\vec{U}_c(x-x_0)$ which fixes the initial value of the protocol of motion as 
$\phi(t_0)=x_0$. Then, the solution \eq{eq:ControlSolutionMultiComponent} for the control signal becomes
\begin{equation}
 \vec{f}(x,t) = \vec{\mathcal{B}}^{-1}(\displaystyle -\dot{\phi}(t)\vec{U}_c'(x-\phi(t)) 
  - \vec{\mathcal{D}}\vec{U}_c''(x-\phi(t)) + \vec{R}(\vec{U}_c(x-\phi(t))),
\end{equation}
with $\dot{\phi}(t)$ denoting the derivative of $\phi(t)$ with respect to time $t$.
Using \eq{eq:ProfileEquation} to eliminate the non-linear reaction kinetics $\vec{R}$, we finally obtain the following analytical expression 
for the control signal
\begin{equation}
 \vec{f}(x,t) =  (\displaystyle c-\dot{\phi}(t))\vec{\mathcal{B}}^{-1}\vec{U}_c'(x-\phi(t)) 
  =: \vec{f}_\mathrm{an}.\label{eq:PositionControlSolution}
\end{equation}
Remarkably, any reference to the reaction function $\vec{R}$ drops out from the expression for the control. 
This is of great advantage for applications without or only incomplete knowledge of the underlying reaction kinetics $\vec{R}$. 
The method is applicable as long as the propagation velocity $c$ is known and the uncontrolled 
wave profile $\vec{U}_c$ can be measured with sufficient accuracy to calculate the derivative $\vec{U}_c'$.

% A general problem of position control, being an open loop control, is its possible inherent instability. 
% The controlled trajectory obtained by numerical or analytical techniques might be 
% unstable against perturbations of the initial conditions, and must be stabilized by an additional feedback control. 
Being an open loop control, a general problem of the proposed position control is its possible inherent instability against perturbations of 
the initial conditions as well as other data uncertainty. However, assuming protocol velocities $\dot{\phi}(t)$ close to 
the uncontrolled velocity $c$, $\dot{\phi}\sim c$, the control signal \eq{eq:PositionControlSolution} is small in amplitude and 
enforces a wave profile which is relatively close to the uncontrolled TW profile $\vec{U}_c$. Since the uncontrolled TW is presumed to be stable, the 
controlled TW might benefit from that and a stable open loop control is expected. This expectation is confirmed numerically 
for a variety of controlled RD systems \cite{loeber_engel2014}, and also analytically in Ref. \cite{lober2014stability}.

\medskip

\noindent Despite the advantages of our analytical solution stated above, there are limits for it as well. The restriction to fully actuated systems, i.e.,
systems for which $\vec{\mathcal{B}}^{-1}$ exists, is not always practical. In experiments with RD systems, the number of state
components is usually much larger than one while the number of control signals is often restricted to one or two. Thus, the question arises if 
the approach can be extended to underactuated systems with a number of independent control signals smaller than the number of 
state components. This is indeed the case but entails additional assumptions about the desired trajectory. In the context of position control 
of TWs, it leads to a control which is not able to preserve the TW profile for all state components, see Ref.~\cite{loeber_engel2014}. The 
general case is discussed in the thesis \cite{lober2015thesis} and is not part of this paper.

Moreover, in applications it is often necessary to impose inequality constraints in form of upper and lower bounds on the control. 
For example, the intensity of a heat source deployed as control is bounded by technical reasons. Even worse, if the control is 
the temperature itself it is impossible to attain negative values. Since the control signal $\vec{f}_\mathrm{an}$ for position control 
is proportional to the slope of the controlled wave profile $\vec{U}_c'$, the magnitude of the applied control may locally attain 
non-realizable values. In our analytic approach no bounds for the control signal are
imposed. The control signal $\vec{f}$ as given by \eq{eq:ControlSolutionMultiComponent} is optimal only in case of a
vanishing Tikhonov-regularization parameter $\nu=0$, cf. \eq{eq:DistdistRegularized}.
Moreover, desired trajectories $\vec{u}_d$ which do not comply with initial as
well as boundary conditions or are non-smooth might be requested. Lastly, the control signal $\vec{f}$
cannot be used in systems where only a restricted region of the spatial domain $\Omega$
is accessible by control. While all these cases cannot be treated within the analytical approach proposed here,
optimal control can deal with many of these complications.

\section{Optimal Control}\label{S2}
\index{Optimal control,FitzHugh-Nagumo}
In the following, we recall the optimal control problem and sketch the most important analytical results to provide the optimality system.
\subsection{The Control Problem}
For simplicity, we state the optimal control problem explicitly for the FitzHugh-Nagumo system \cite{nagumo62,fitz61}. 
The FitzHugh-Nagumo system is a two-component model $\vec{u} = (u,v)^T$ for an activator $u$ and an inhibitor $v$,
\begin{equation}
\begin{array}{rcl}
\displaystyle \partial_t u(\vec{x},t) - \Delta u(\vec{x},t) + R(u(\vec{x},t)) + \alpha\, v(\vec{x},t) & = &f(\vec{x},t), \\
\displaystyle \partial_t v(\vec{x},t) + \beta \, v(\vec{x},t) - \gamma \,u(\vec{x},t) + \delta & = & 0,
\end{array}
\label{E1.1}
\end{equation}
in a bounded Lipschitz-domain $\Omega \subset \R^N$ of dimension $1 \leq N \leq 3$. 
Since the single-component control $f$ appears solely on the right-hand side of the first equation, this system is underactuated. 
Allowing a control in the second equation is fairly analogous. The kinetic parameters $\alpha$, $\beta$, $\gamma$, and $\delta$ are real 
numbers with $\beta \geq 0$. Moreover, the reaction kinetics is given by the nonlinear function $R(u)=u(u-a)(u-1)$ for 
$0 \leq a \leq 1$. 
Note that the equation for the activator $u$ decouples from the equation for the inhibitor $v$ for $\alpha = 0$, cf. \eqs{E1.1},
resulting in the Schl\"ogl model \cite{schloegl72,Zeldovich1938}, sometimes also called the Nagumo model. 
We assume homogeneous Neumann-boundary conditions for the activator $u$ and $u(\vec{x},0) = u_0(\vec{x})$, 
$v(\vec{x},0) = v_0(\vec{x})$ are given initial states belonging to $L^\infty(\Omega)$, i.e., they are bounded. 

The aim of our control problem is the tracking of desired trajectories $\vec{u}_d = (u_d,v_d)^T$ in the space-time cylinder $Q$
and to reach desired terminal states $\vec{u}_T= (u_T,v_T)^T$ at the final time $T$. In contrast to the analytic approach from Sect. \ref{S1}, 
these desired trajectories are neither assumed to be smooth nor compatible with the given initial data or boundary conditions. 
For simplicity, we assume their boundedness, i.e., $(u_d,v_d)^T \in \left(L^\infty(Q)\right)^2$ 
and $(u_T, v_T)^T \in \left(L^\infty(\Omega)\right)^2$.
The goal of reaching the desired states is expressed as the minimization of the objective functional
\begin{equation}\begin{split}
J(u,v,f) & = \frac{1}{2}\left(c_T^U\|u(\,\cdot\,,T)-u_T\|_\ltwo^2 + c_T^V\|v(\,\cdot\,,T)-v_T\|_\ltwo^2\right)\\
	 & + \frac{1}{2}\left(c_d^U\|u-u_d\|_\ltwoq^2 + c_d^V\|v-v_d\|_\ltwoq^2\right) + \frac{\nu}{2}\|f\|_\ltwoq^2.
\end{split}\label{eq:JFHN}\end{equation}
This functional is slightly more general than the one given by \eq{eq:Distdist} because it also takes into account the
terminal states. We emphasize that the given non-negative coefficients $c_d^U,\, c_d^V, \, c_T^U$, and $c_T^V$ 
can also be chosen as functions depending on space and time. In some applications, this turns out to be very useful 
\cite{casas_ryll_troeltzsch2014}. The control signals can be taken out of the set of admissible controls
\begin{equation}
\Uad = \{f \in L^\infty(Q) : f_a \le f(\vec{x},t) \le f_b,\ \mbox{ for } (\vec{x},t) \in Q\}.\label{E1.2}
\end{equation}
The bounds $-\infty < f_a < f_b < \infty$ model the technical capacities for generating controls. 
\medskip

\noindent Under the previous assumptions, the controlled RD equations \eqref{E1.1} 
have a unique weak solution denoted by $(u_f,v_f)^T$ for a given control $f \in \Uad$. This solution is bounded, i.e., 
$u_f,\, v_f \in L^\infty(Q)$, cf. \cite{casas_ryll_troeltzsch2014}. If the initial data $(u_0, v_0)^T$ are continuous then $u_f$ and $v_f$ are continuous on 
$\bar \Omega \times [0,T]$ with $\bar \Omega=\Omega \cup \partial \Omega$ as well. Moreover, the control-to-state mapping  
$G:=f \mapsto (u_f,v_f)^T$ is twice continuously (Fr\`echet-) differentiable. A proof can be found in 
Ref.~\cite[Theorem 2.1,Corollary 2.1, and 
Theorem 2.2]{casas_ryll_troeltzsch2014}. Expressed in terms of the solution $(u_f,v_f)^T$, the value of the objective functional depends
only on $f$, $J(u,v,f) =J(u_f,v_f,f) =: F(f)$, and the optimal control problem can be formulated in a condensed form as
\begin{align}
  \Pb\qquad {\rm Min} \ F(f), \quad f \in \Uad . \label{eq:OptProbP}
\end{align}

Referring to \cite[Theorem 3.1]{casas_ryll_troeltzsch2014}, we know that the control problem \Pb has at least one (optimal)
solution $\bar f$ for all $\nu \ge 0$ . To determine this solution numerically, we need the first and second-order derivatives of the 
objective function $F$. Since the mapping $f \mapsto (u,v)^T$ is twice continuously differentiable, so is 
$F:L^p(Q) \longrightarrow \mathbb{R}$. Its first derivative $F'(f)$ in the direction $h\in L^p(Q)$ can be computed as follows
\begin{equation}
F'(f)h = \intQb{(\varphi_f + \nu f)h},
\label{E1.5}
\end{equation}
where $\varphi_f$ denotes the first component of the so-called adjoint state $(\varphi_f,\psi_f)$. It solves a linearized FitzHugh-Nagumo system, backwards in time,
\begin{equation}
\begin{array}{rcl}
\displaystyle -\partial_t \varphi_f - \Delta \varphi_f + R^\prime(u_f)\varphi_f - \gamma\, \psi_f & = &c_d^U(u_f-u_d), \\
\displaystyle -\partial_t \psi_f + \beta \, \psi_f + \alpha \,\varphi_f & = & c_d^V(v_f-v_d),
\label{AdjointSystem}
\end{array}
\end{equation}
with homogeneous Neumann-boundary and terminal conditions 
$\varphi_f(\vec{x},T) = c_T^U(u_f(\vec{x},T)-u_T(\vec{x}))$ and 
$\psi_f(\vec{x},T) = c_T^V(v_f(\vec{x},T)-v_T(\vec{x}))$ in $\Omega$.

This first derivative is used in numerical methods of gradient type. Higher order methods of Newton type need 
also the second derivative $F^{\prime\prime}(f)$.
It reads
\begin{equation}\begin{split}
& F''(f)h^2 = \intOmb{c_T^U\eta_{h}(\vec{x},T)^2 + c_T^V\zeta_{h}(\vec{x},T)^2}\\
& \qquad + \intQb{[c_d^U - R''(u_f)\varphi_f]\eta_{h}^2+ c_d^V\zeta_{h}^2} + \nu\intQb{h^2},
\end{split}\label{E1.6}\end{equation}
in a single direction $h \in L^p(Q)$. In this expression, the adjoint state $(\eta_{h},\zeta_{h}) := G^\prime(f)h$ denotes the solution of a linearized FitzHugh-Nagumo system similar to \eq{AdjointSystem}, see Ref.~\cite[Theorem 2.2]{casas_ryll_troeltzsch2014} for more information. 

%%%%%%%%%%%%%%%%%%%%%%%
\subsection{First-Order Optimality Conditions}
%%%%%%%%%%%%%%%%%%%%%%%

We emphasize that the control problem \Pb is not necessarily convex. 
Although the objective function $J(u,v,f)$ is convex, in general the nonlinearity of the mapping $f \mapsto (u_f,v_f)^T$ will lead 
to a non-convex function, $F$. Therefore, (P) is a problem of non-convex optimization, possibly leading to several 
local minima instead of a single global minimum.

As in standard calculus, we invoke first-order necessary optimality conditions to find a (locally) optimal control $f$, denoted by $\bar f$.
In the case of unconstrained control, i.e., $\Uad := L^p(Q)$, the first derivative of $F$ must 
be zero, $F^\prime(\bar f) = 0$. Computationally, this condition is better expressed in the weak formulation
\begin{equation}
   F^\prime(\bar f)f = \intQb{(\bar\varphi + \nu \bar f)f} = 0,\quad \forall f \in L^p(Q),\label{E1.3}
\end{equation}
where $\bar\varphi$ denotes the first component of the adjoint state associated with $\bar f$.
If $\bar f$ is not locally optimal, one finds a descent direction $d$ such that 
$F^\prime(\bar f)d < 0$. This is used for methods of gradient type.

If the restrictions $\Uad$ are given by \eq{E1.2}, then \eq{E1.3} does not hold true in general. Instead, the variational inequality
\begin{equation}
  F^\prime(\bar f)(f - \bar f) = \intQb{(\bar\varphi + \nu \bar f)(f - \bar f)} \geq 0,\quad \forall f \in \Uad \cap U(\bar f),\label{E1.4}
\end{equation}
must be fulfilled, cf. \cite{tro10book}. Here $U(\bar f) \subset L^p(Q)$ denotes a neighborhood of $\bar f$. 
Roughly speaking, it says that in a local minimum we cannot find an admissible direction of descent. 
A gradient method would stop in such a point. A pointwise discussion of \eq{E1.4} leads to the following identity:
\begin{equation}
\displaystyle \bar f(\vec{x},t) = \mbox{\rm
Proj}_{[f_a,f_b]}\left(-\frac{1}{\nu}[\bar\varphi(\vec{x},t)]\right), \text{ for } \nu > 0.
\label{E2.35}
\end{equation}
Here, $\mbox{\rm Proj}_{[f_a,f_b]}(x) = \min\{\max\{f_a , x\}, f_b\}$ denotes the projection to the interval $[f_a,f_b]$ such
that $\bar f(\vec{x},t)$ belongs to the set of admissible controls $\Uad$ defined in \eq{E1.2}.
According to \eq{E2.35}, as long as $\bar\varphi$ does not vanish, a decreasing value 
of $\nu\geq0$ yields an optimal control growing in amplitude until it attains its bounds $f_a$ or $f_b$, respectively.
Thus, the variational inequality \eq{E1.4} leads to so-called bang-bang-controls \cite{tro10book} for $\nu=0$ and $\bar\varphi \neq 0$.
These are control signals which attain its maximally or minimally possible values for all times and everywhere in the spatial domain $\Omega$.
A notable exception is the case of exactly realizable desired trajectories and $\nu=0$, already discussed in Sect. \ref{S1}. In this case, it can be shown that
$\bar\varphi$ vanishes \cite{lober2015thesis} and \eq{E2.35} cannot be used to determine the control signal $\bar f$. 

Numerically, solutions to optimal control are obtained by solving the controlled RD system \eq{E1.1} and the adjoint system, \eq{AdjointSystem},
such that the last identity, \eq{E2.35}, is fulfilled. In numerical computations with very large or even missing bounds $f_a,\,f_b$, 
\eq{E2.35} becomes ill-conditioned if $\nu$ is close to zero. This might lead to large roundoff errors in the computation of 
the control signal and can affect the stability of numerical optimal control algorithms.

\subsection{Example 1: Analytical and Optimal Position Control} \label{S2-3}
\index{Schlögl model,Traveling fronts}
In 1972, Schl\"ogl discussed the auto-catalytic trimolecular RD scheme \cite{schloegl72,Zeldovich1938} as a prototype of a 
non-equilibrium first order phase transition. The reaction kinetics $R$ for the chemical with concentration 
$u(x,t)$ is cubic and can be casted into this dimensional form $R(u) = u(u-a)(u-1)$. 
The associated controlled RD equation reads 
\begin{align*}
\partial_t u- \partial_x^2 u + u\bracket{u-a}\bracket{u-1}=\,f(x,t),\quad 0<a<1, 
\end{align*}
in one spatial dimension, $x \in \R$. A linear stability analysis of the uncontrolled system reveals that $u=0$ and $u=1$ are spatially 
homogeneous stable steady states (HSS) while $u=a$ is an unstable homogeneous steady state. 
% Therefore, the Schl\"ogl model describes a bistable chemical reaction in a certain parameter range. 
In an infinite one-dimensional domain, the Schl\"ogl model possesses a stable traveling front solution whose profile is given by
\begin{align}
U_c(\xi) = 1/\left(1+\exp \left(\xi/\sqrt{2}\right)\right),\label{eq:SchloeglFrontSolution}
\end{align}
in the frame of reference $\xi=x-c\,t$ co-moving with front velocity $c$. This front solution establishes a heteroclinic connection between the two stable HSS 
for $\xi\rightarrow\pm\infty$ and travels with a velocity $c = \bracket{1-2\,a}/\sqrt{2}$ from the left to the right. 
% Noteworthy, the front velocity depends on the excitation thres\-hold $a$ while the front profile is independent of $a$. 
% This is a peculiarity of the Schl\"ogl model.

\begin{figure}[b]
  \centering
  \subfloat[\label{F:1a}]{\includegraphics[width=0.333\textwidth]{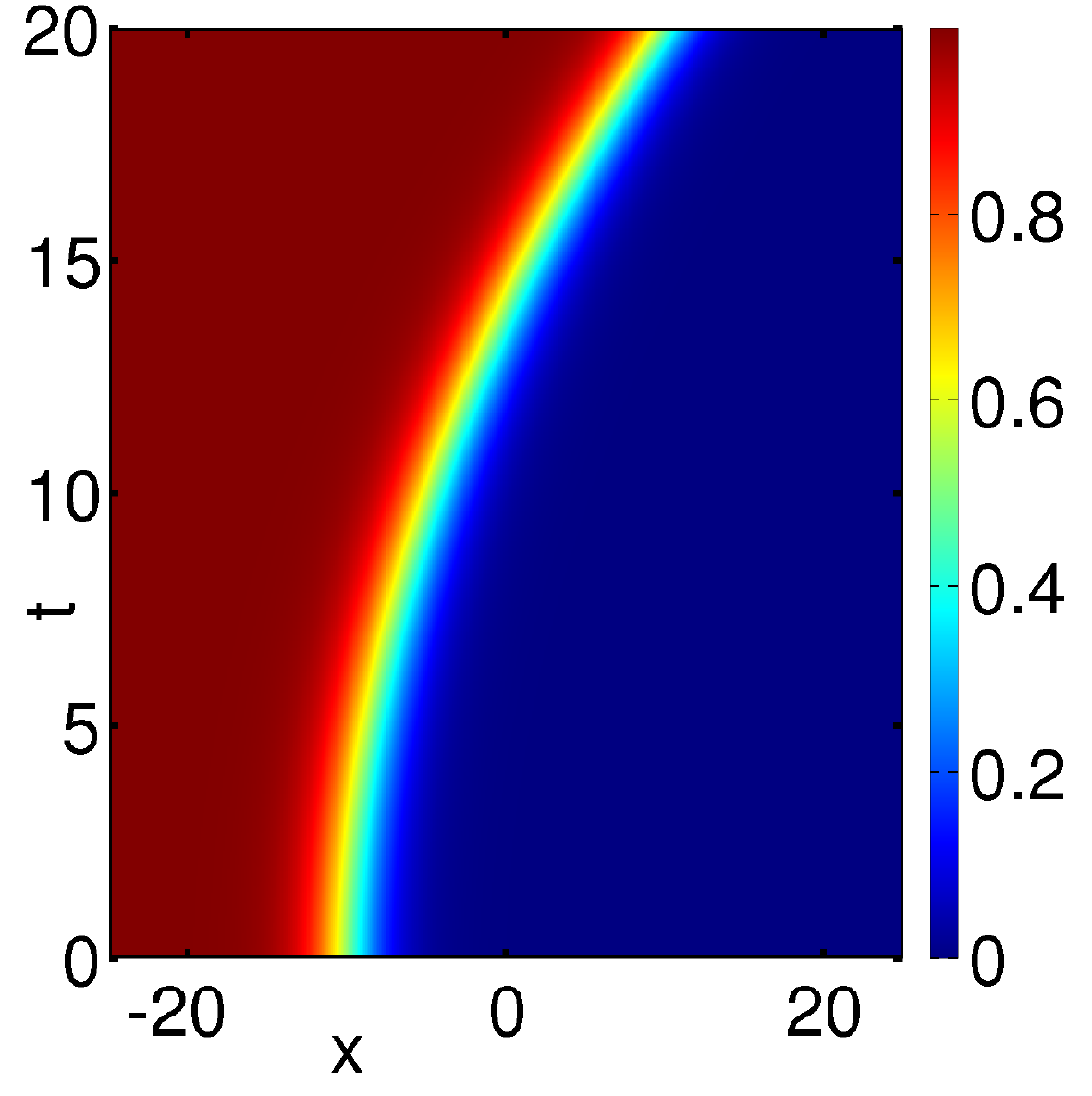}}
  \subfloat[\label{F:1b}]{\includegraphics[width=0.333\textwidth]{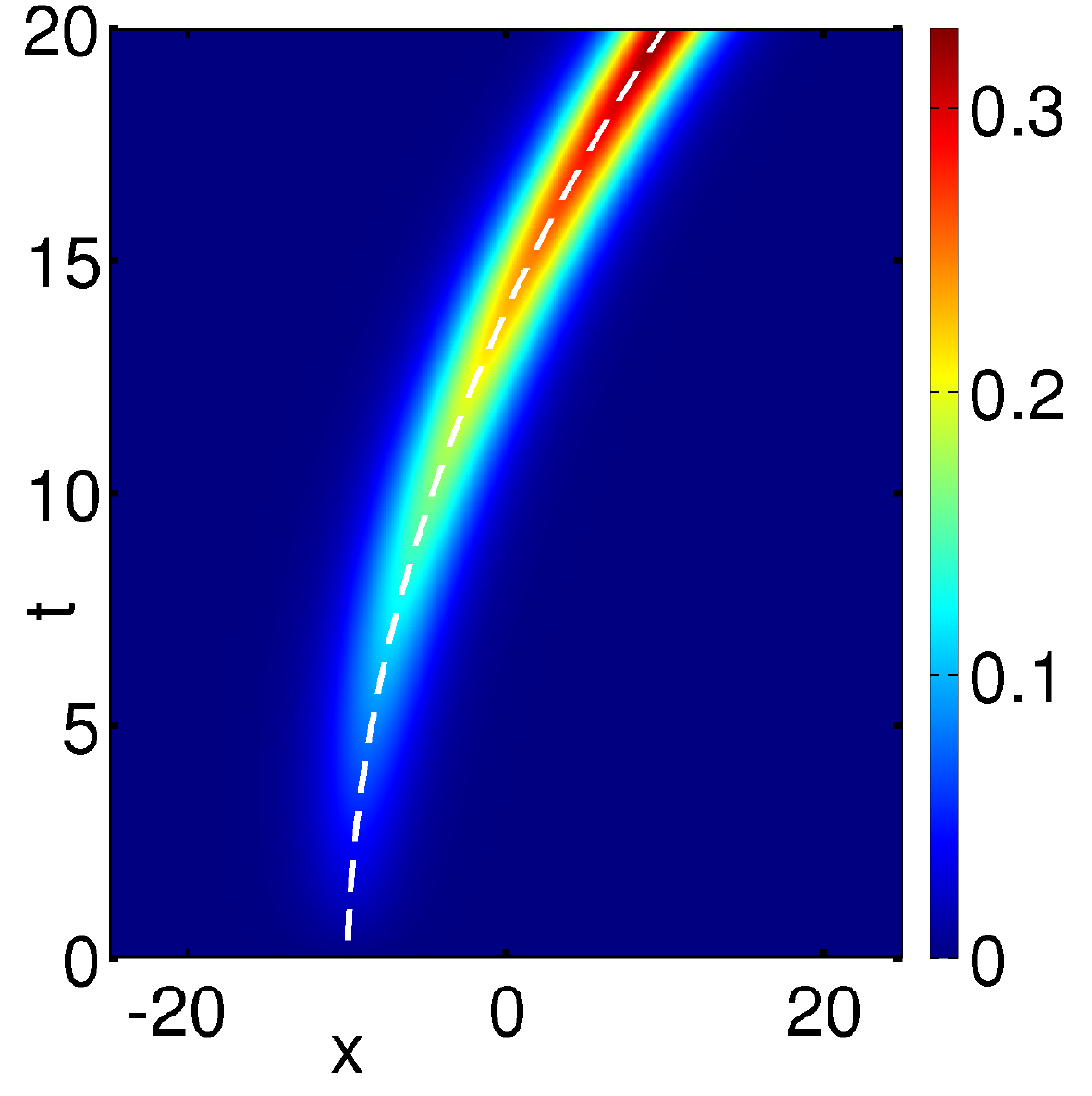}}
  \subfloat[\label{F:1c}]{\includegraphics[width=0.333\textwidth]{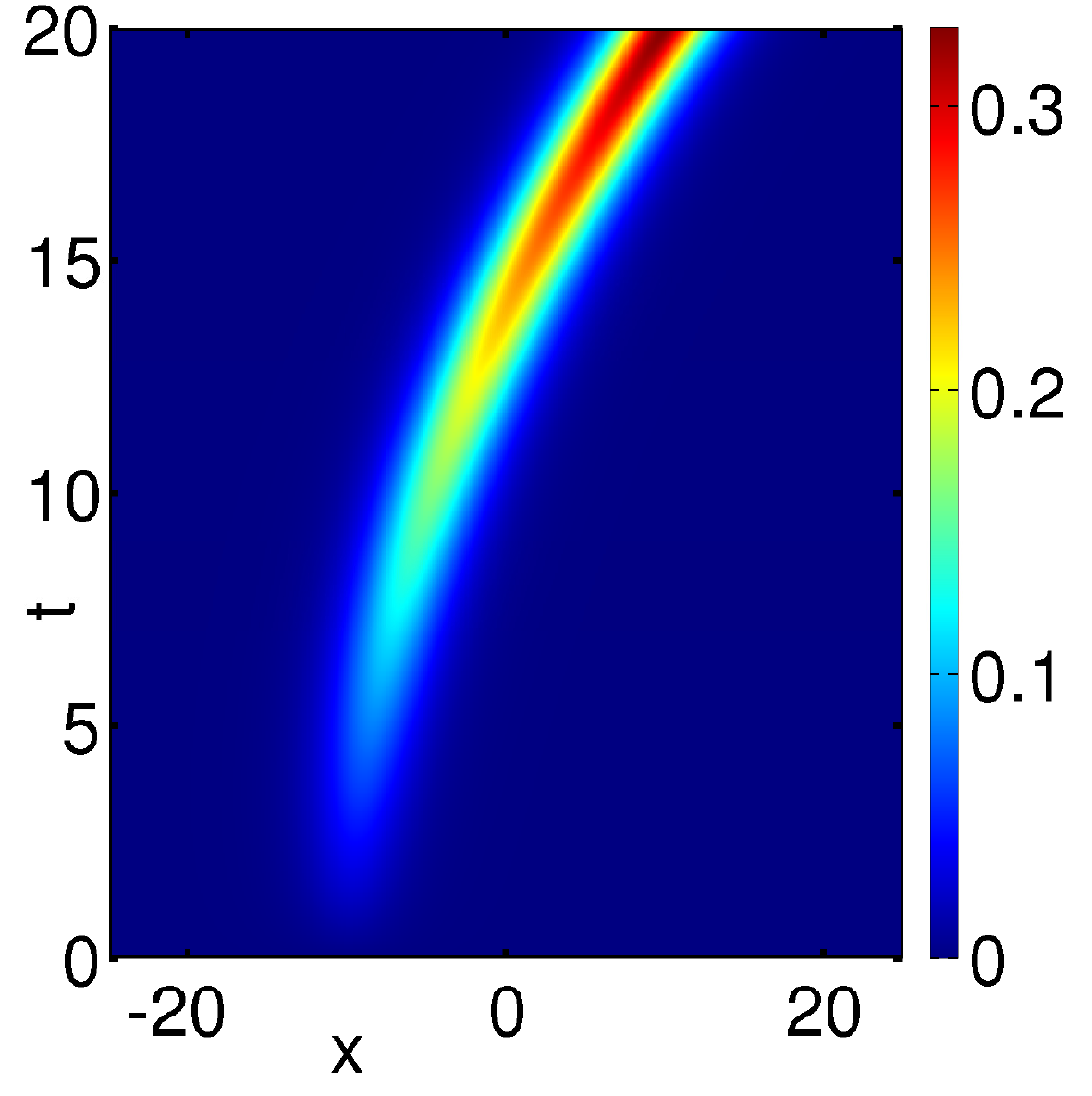}}
  \caption{\textbf{(a)} Space-time plot of the desired trajectory $u_d(x,t) = U_c(x-\phi(t))$ with the protocol of motion $\phi(t)$ as given in \eq{eq:protocol_Schlogl}, \textbf{(b)} analytic position control signal $f_\text{an}(x,t)$, \eq{eq:PositionControlSolution}, and \textbf{(c)} numerically obtained optimal control $\bar f$ for Tikhonov regularization parameter $\nu = 10^{-5}$ are presented. The magnitude of the control signal is color-coded. In the center panel \textbf{(b)}, the dashed line represents $\phi(t)$. The remaining parameter values are $a=9/20$, $T=20$, and $c_d^U = c_T^U = 1$.}
  \label{F:1}
\end{figure}

As an example, we aim to accelerate a traveling front according to the following protocol of motion
\begin{align}
 \phi(t) = -10 + c \,t + \displaystyle\frac{10-1/\sqrt{2}}{200} \, t^2, \label{eq:protocol_Schlogl}
\end{align}
while keeping the front profile as close as possible to the uncontrolled one. In other words, our desired trajectory reads
$u_d(x,t) = U_c(x-\phi(t))$, and consequently the initial condition of both the controlled and the desired trajectory are $u_0(x)=u_d(x,0)=U_c(x+10)$. 
In our numerical simulations, we set $T=20$ for the terminal time $T$, $\Omega = (-25,25)$ for the spatial domain, and the threshold parameter 
is kept fixed at $a=9/20$. Additionally, we choose the terminal state to be equal to the desired trajectory, 
$u_T(x) = u_d(x,T)$, and set the remaining weighting coefficients to unity, $c_d^U = c_T^U = 1$, in the optimal control problem. 
The space-time plot of the desired trajectory $u_d$ is presented in Fig. \ref{F:1a} 
for the protocol of motion $\phi(t)$ given by \eq{eq:protocol_Schlogl}.

Below, we compare the numerically obtained solution to the optimal control problem (P) with the analytical solution
from Sect.~\ref{S1} for the Schl\"ogl model. The Schlögl model arises from Eq. (12) by setting $\alpha=0$ and ignoring
the inhibitor variable $v$. Consequently, all weighting coefficients associated with the inhibitor trajectory 
are set to zero, $c_d^V = c_T^V = 0$, in the functional $J$, \eq{eq:JFHN}. 

Fig.~\ref{F:1b} depicts the solution for the analytical position control $f_\text{an}$ which is valid for a vanishing 
Tikhonov regularization parameter $\nu = 0$. The numerically obtained optimal control $\bar{f}$ for $\nu = 10^{-5}$, shown in Fig.~\ref{F:1c}, 
does not differ visually from the analytic one. Both are located at the front position where the slope is maximal, 
$\vec{u}_d=0.5$ (dashed line in Fig.~\ref{F:1b}), and their magnitudes grow proportional to $\dot{\phi}(t)$. For a quantitative 
comparison, we compute the distance between analytical and optimal control signal $\|\bar f - f_\text{an}\|_2$ in the sense of $L^2(Q)$, 
\eq{eq:L2norm}, and normalized it by the size of the space-time-cylinder $|Q| = T\,|\Omega|$
\begin{equation}
 \|h\|_2 := \frac{1}{|Q|}\|h\|_{L^2(Q)}.\label{E:norm}
\end{equation}
The top row of Tab.~\ref{T1} displays the distance $\|\bar f - f_\text{an}\|_2$ as a 
function of the regularization parameter $\nu$. 
Even for a large value of $\nu=1$, the distance is less than $5\times 10^{-4}$. Decreasing the value of $\nu$ results in a shrinking distance $\|\bar f - f_\text{an}\|_2$ until it saturates at $\simeq 8\times 10^{-6}$. The saturation is due to numerical and systematic errors. Numerical computations are affected by errors arising in the discretization of the spatio-temporal domain and the amplification of roundoff errors by the ill-conditioned expression for the control, \eq{E2.35}. A systematic error arises because the optimal control is computed for a bounded interval $\Omega=(-25,25)$ with homogeneous Neumann-boundary 
conditions while the analytical result is valid only for an infinite domain.

\begin{table}[b]\centering\small
  \caption{The distance $\|\bar f - f_\text{an}\|_2$ between the analytical control signal 
  $f_\text{an}$, valid for $\nu=0$, and the optimal control $\bar f$ obtained numerically for finite 
  $\nu>0$  decreases with decreasing values of $\nu$ (top row). Similarly, the optimally controlled state trajectory 
  $u_f$ approaches the desired trajectory $u_d$, measured by $\|u_f - u_d\|_2$, for smaller values of $\nu$ (bottom row).}
  \begin{tabular}{|c||c|c|c|c|c|c|c|}
  \hline
  $\nu$ & 1 & E-1 & E-2 & E-3 & E-4 & E-5 & E-6\\
  \hline
  $\|\bar f - f_\text{an}\|_2$ & 4.57E-4 & 1.14E-4 & 2.50E-5 & 1.01E-5 & 8.40E-6 & 8.30E-6 & 8.29E-6\\[.5ex]
  $\|u_f - u_d\|_2$           & 4.77E-4 & 7.49E-5 & 8.34E-6 & 8.52E-7 & 8.55E-8 & 8.56E-9 & 8.56E-10 \\
  \hline
  \end{tabular}
  \label{T1}
\end{table}

Another interesting question is how close the controlled state $u_f$ approaches the desired trajectory $u_d$. The bottom row of 
Tab.~\ref{T1} shows the distance between the optimal controlled state trajectory $u_f$ and the desired trajectory for different values of 
$\nu$. Similarly as for the control signal, the difference lessens with decreasing values of $\nu$. Note that the value does not saturate 
and becomes much smaller than the corresponding value for the difference between control signals. 
Here, no discretization errors arise because a discretized version of the desired trajectory is used as the target distribution. Nevertheless, 
systematic errors arise because neither the initial and final desired state nor the desired trajectory obey Neumann-boundary conditions.
This results in an optimal control signal exhibiting bumps close to the domain boundaries. However, the violation of boundary conditions 
can be reduced by specifically designed protocols of motion. The further the protocol of motion keeps the controlled front 
away from any domain boundary the smaller is the violation of homogeneous Neumann-boundary conditions since the derivatives of traveling front solution 
\eq{eq:SchloeglFrontSolution} decay exponentially for large $|x|$. An alternative way of rigorously avoiding artifacts due to the violation of boundary conditions is the introduction of additional control terms acting on the domain boundaries, see Ref.~\cite{lober2014shaping}.
%In our numerical example, the maximal violation of the Neumann-boundary condition is approximately $1.75 \times 10^{-5}$.

For the example discussed above the numerical optimal control $\bar f$ for $\nu>0$ is computed with a Newton-Raphson-type root finding algorithm. 
This iterative algorithm relies on an initial guess for the control signal, which is often chosen to be random or uniform in space. The closer
the initial guess is to the final solution, the fewer steps are necessary for the Newton-Raphson method to converge on the final solution. 
The similarity of the numerical and analytical control solution, see Fig.~\ref{F:1} and Tab.~\ref{T1}, motivates the utilization of the analytical result 
$f_\text{an}$ as an initial guess in numerical algorithms.
Even for a simple single component RD system defined on a relatively small spatio-temporal domain $Q$ as discussed in this section, the 
computational speedup is substantial. The algorithm requires only $2/3$ of the computation time compared to random or uniform 
starting values for the control. In particular, we expect even larger speedups for simulations with larger domain sizes.
\FloatBarrier

%%%%%%%%%%%%%%%%%%%
\section{Sparse Optimal Control}\label{S3}
\index{Sparse optimal control}
%%%%%%%%%%%%%%%%%%%%

In applications, it might be desirable to have localized controls acting only in some sub-areas of the domain. 
So-called sparse optimal controls provide such solutions without any a priori knowledge of these sub-areas. 
They result in a natural way because the control has the best influence in these regions to achieve a certain objective functional 
to be minimized.

For inverse problems, it has been observed that the use of an $L^1$-term in addition to the $L^2$-regularization leads to sparsity \cite{DaubechiesEA04,Vogel02,ChanTai03}. The idea to use the $L^1$-term goes back to Ref.~\cite{RudinEA92}.

To our knowledge, sparse optimal controls were first discussed in the context of optimal control in Ref. \cite{stadler2009}. 
In that paper, an elliptic linear model was discussed. Several publications followed, investigating semi-linear elliptic equations, 
parabolic linear, and parabolic semi-linear equations; we refer for instance to Refs. 
\cite{wachsmuth_wachsmuth2011,CHW2012a,casas_troeltzsch_sparse_state2013} 
among others.

In this section, we follow the lines of Refs.~\cite{casas_ryll_troeltzsch2014, casas_ryll_troeltzsch2014b} and 
recall the most important results for the sparse optimal control of the Schl\"ogl-model and the FitzHugh-Nagumo equation. 

\subsection{The Control Problem}

In optimal control, sparsity is obtained by extending the objective functional $J$ by a multiple of $j(f) := \|f\|_{L^1(Q)}$, 
the $L^1$-norm of the control $f$. Therefore, recalling that $J(u_f,v_f,f) =: \mathcal{F}(f)$, we consider the problem
\[
 \Pbsp \qquad {\rm Min} \ \mathcal{F}(f) + \kappa\, j(f), \quad f \in \Uad .
\]
for $\kappa > 0$. The first part $F$ of the objective functional is differentiable while the $L^1$-part is not. 

Our goal is not only to derive first-order optimality conditions as in the previous section but also to observe the 
behavior of the optimal solutions for increasing $\kappa$ and $\nu$ is tending to zero. 
For that task we also need to introduce second-order optimality conditions.

As before, there exists at least one locally optimal solution $f$ to the problem \Pbsp, denoted by $\bar f$. 
We refer to Ref. \cite[Theorem 3.1]{casas_ryll_troeltzsch2014} for more details. 
While $\mathcal{F}$ is twice continuously differentiable, the second part $j(f)$ is only Lipschitz convex
but not differentiable. For that reason, we need the so-called subdifferential of $j(f)$.
By subdifferential calculus and using directional derivatives of $j(f)$, we are able to derive necessary optimality conditions.

%%%%%%%%%%%%%%%%%%%%%%%%
\subsection{First-Order Optimality Conditions}
%%%%%%%%%%%%%%%%%%%%%%%%

We recall some results from Refs. \cite{casas_ryll_troeltzsch2014, casas_ryll_troeltzsch2014b}.
Due to the presence of $j(f)$ in the objective functional, there exists a $\bar\lambda \in \partial j(\bar f)$ such that
the variational inequality \eq{E1.4} changes to
\begin{equation}
\intQb{(\bar\varphi + \nu\bar f + \kappa\bar\lambda)(f - \bar f)} \ge 0,\quad \forall f \in \Uad \cap U(\bar f).
\label{E2.3}
\end{equation}
For the problem \Pbsp, a detailed and extensive discussion of the first-order necessary optimality condition leads to very 
interesting conclusions, namely
\begin{align}
\displaystyle \bar f(\vec{x},t) & = 0, \text{ if and only if } |\bar\varphi(\vec{x},t)| \le \kappa,\label{E2.4a}\\
\displaystyle \bar f(\vec{x},t) & 
  = \mbox{\rm Proj}_{[f_a,f_b]}\left(-\frac{1}{\nu}[\bar\varphi(\vec{x},t)+\kappa\bar\lambda(\vec{x},t)]\right),\label{E2.4b}\\
\displaystyle\bar\lambda(\vec{x},t) & = \mbox{\rm Proj}_{[-1,+1]}\left(-\frac{1}{\kappa}\bar\varphi(\vec{x},t)\right),\label{E2.4c}
\end{align}
if $\nu > 0$. We refer to Refs.~\cite[Corollary 3.2]{CHW2012a} and \cite[Theorem 3.1]{Casas2012} in which the case $\nu = 0$ is discussed as well.

The relation in \eq{E2.4a} leads to the sparsity of the (locally) optimal solution $\bar f$, depending on the sparsity parameter $\kappa$.
In particular, the larger the choice of $\kappa$ is the smaller does the support of $\bar f$
become. To be more precise, there exists a value $\kappa_0 > \infty$ such that for every 
$\kappa \geq \kappa_0$ the only local minimum $\bar f$ is equal to zero. Obviously, this case is ridiculous and thus, one needs
some intuition to find a suitable value $\kappa$. We emphasize that $\bar\lambda$ is unique, see \eq{E2.4c}, which is important 
for numerical calculations.

\subsection{Example 2: Optimal and Sparse Optimal Position Control} \label{SS3.1}

For the numerical computations, we follow the lines of Ref. \cite{casas_ryll_troeltzsch2014} and use a non-linear 
conjugate gradient method. The advantage of using a (conjugate) gradient method lies in the simplicity in its implementation
and in the robustness of the method to errors in the solution process. 
Moreover, it allows to solve the systems \eq{E1.1} and the adjoint system separately. The disadvantage is clearly the fact that it 
might cause a huge amount of iterations to converge, cf. Ref. \cite[section 4]{casas_ryll_troeltzsch2014}.

Hence, we modify our approach by the use of \emph{Model Predictive Control} \cite{Propoi63,CamachoBordans99}. 
The idea is quite simple: Instead of optimizing the 
whole time-horizon, we only take a very small number of time steps, formulate a sub-problem and solve it. Then, the first computed 
time-step of the solution $\bar f$ of this smaller problem is accepted on $[0,t_1]$ and is fixed. A new sub-problem is defined by going 
one time-step further and so on. Although the control gained in this way is only sub-optimal, it leads to a much better 
convergence-behavior in many computations.\medskip

\begin{figure}[ht]
  \centering
  \subfloat[\label{F:2a}]{\includegraphics[width=0.333\textwidth]{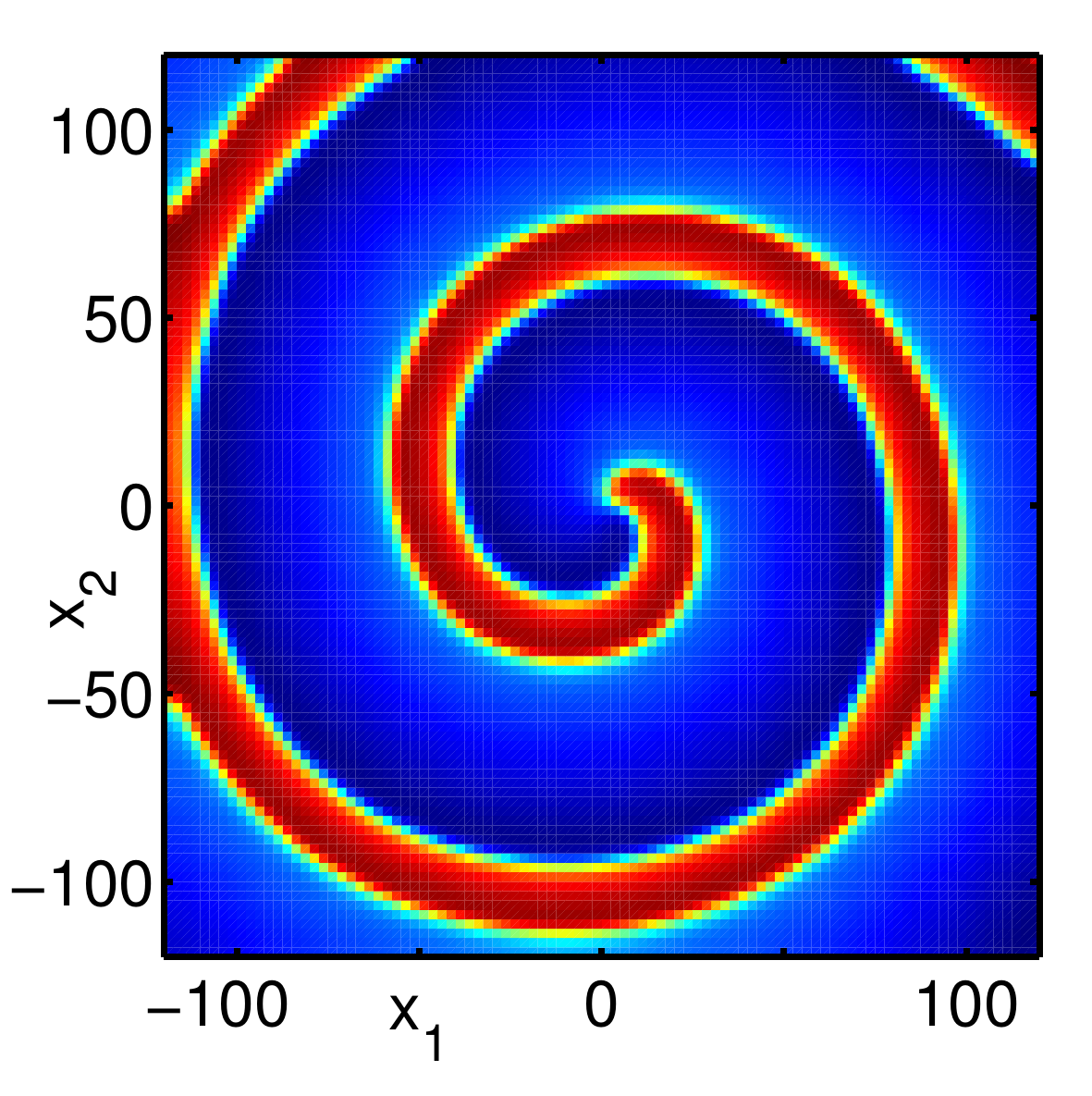}}
  \subfloat[\label{F:2b}]{\includegraphics[width=0.333\textwidth]{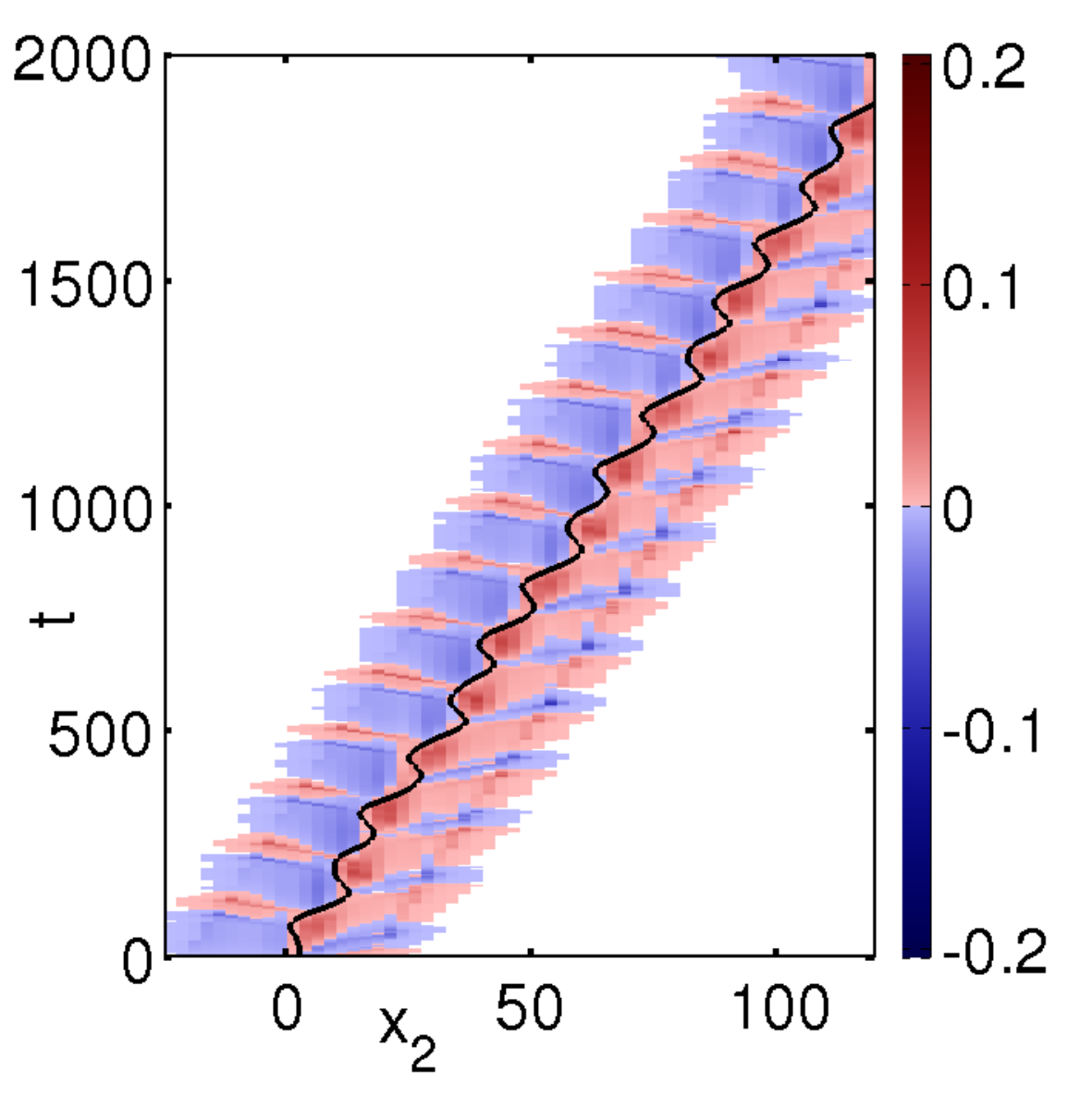}}
  \subfloat[\label{F:2c}]{\includegraphics[width=0.333\textwidth]{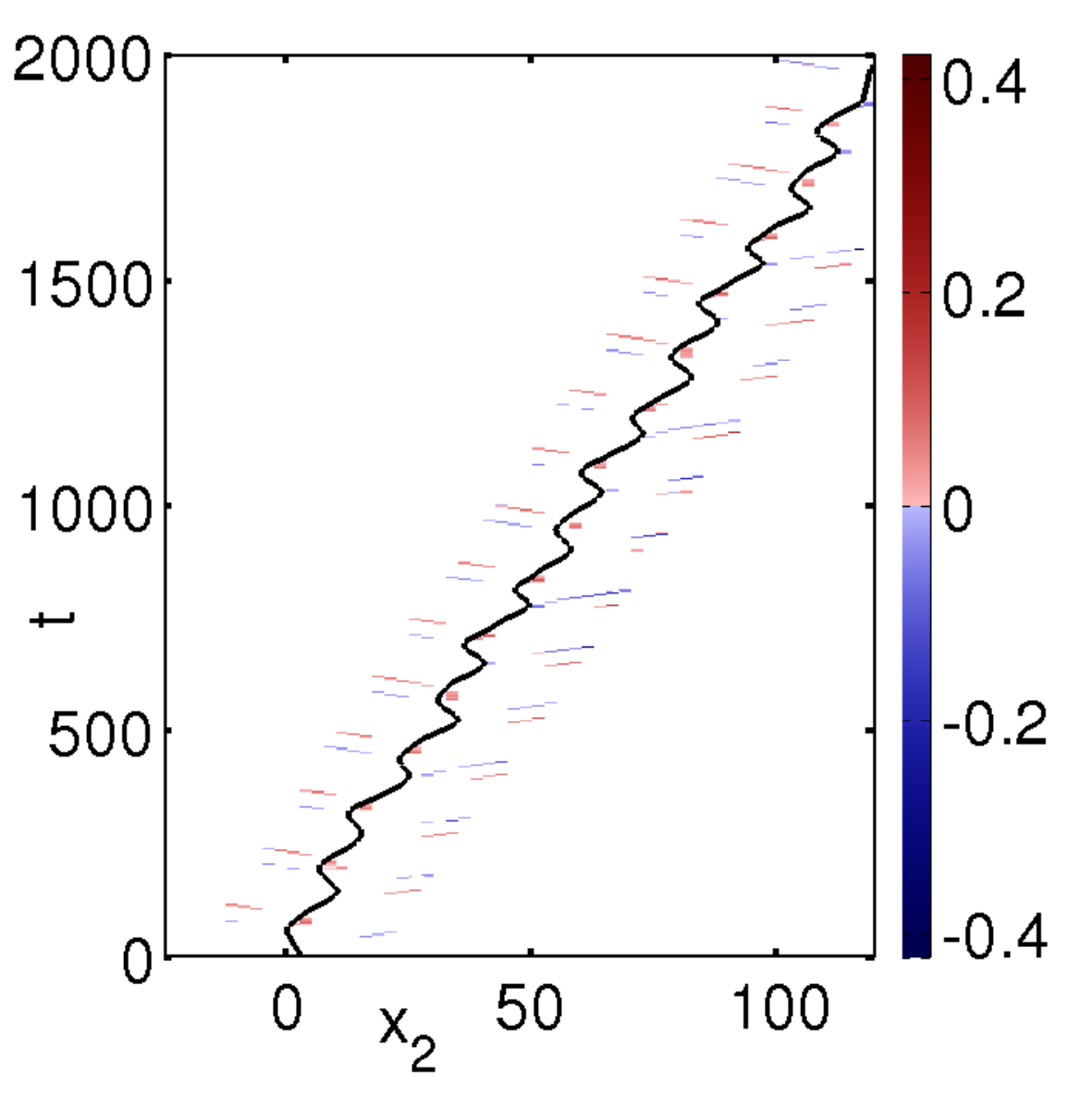}}
  \caption{\textbf{(a)} Spiral wave solution of the activator $u$ to \eq{E1.1} with $\vec{f}=0$. The latter is used as initial state $u_0(\vec{x})=u_\text{nat}(\vec{x}-\vec{\phi}(0),0)$  in the problem \Pbsp. \textbf{(b)}
  Numerically obtained sub-optimal control ($\kappa = 0$) and \textbf{(c)} sparse sub-optimal control solution ($\kappa = 1$), 
  both shown in the $x_2$--$t$--plane for $x_1 = 0$ with associated spiral-tip trajectory (black line). The magnitude of the control signal is color-coded. The remaining system parameters are 
  $a = 0.005$, $\alpha = 1$, $\beta = 0.01$, $\gamma = 0.0075$, and $\delta = 0$. In the optimal control algorithms, 
  we set $\nu = 10^{-6}$, $f_a = -5$, and $f_b = 5$.}
  \label{F:2}
\end{figure}

Next, we revisit the task to extinguish a spiral wave by controlling its tip dynamics such that the whole 
pattern moves out of the spatial domain towards the Neumann boundaries \cite{steinbock1993control,ZykovPhysD2004,schlesner2008efficient}. 
To this goal, following Ex. 6 from Ref. \cite[Section 4]{casas_ryll_troeltzsch2014}, we set the protocol of motion to 
$\vec{\phi}(t) = \left(0,\min\{120, 1/16\,t\}\right)^T$ and $u_d(\vec{x},t) := u_\text{nat}(\vec{x}-\vec{\phi}(t),t)$ 
where $u_\text{nat}$ denotes the naturally develop spiral wave solution of the activator $u$ to \eq{E1.1} for $f = 0$. In our numerical simulation, 
we take only 4 time-steps in each sub-problem of the receding horizon. Moreover, we set the kinetic parameters in the FHN  
model, \eq{E1.1}, to $a = 0.005$, $\alpha = 1$, $\beta = 0.01$, $\gamma = 0.0075$, and $\delta = 0$. Further, we fix the 
simulation domain $\Omega = (-120,120)\times(-120,120)$, the terminal time $T = 2000$, $\nu = 10^{-6}$ as Tikhonov parameter, 
and $f_a = -5$ and $f_b = 5$ as bounds 
for the control, respectively. As initial states $(u_0,v_0)^T$ a naturally developed spiral wave whose core is located at $(0,0)$ 
is used; $u_0$ is presented in Fig.~\ref{F:2a}.

In addition, an observation-function $c_d^U \in L^\infty(Q)$ instead of the constant factor $c_d^U \in \R$ is used with a
support restricted to the area close to the desired spiral-tip. To be more precise, $c_d^U(\vec{x},t) = 1$ holds only in the area defined 
by all $(\vec{x},t) \in Q$ such that $|\vec{x}-\bar{\vec{x}}(t)| \leq 20$ and vanishes identically otherwise. The other coefficients 
$c_d^V$, $c_T^U$, and $c_T^V$ are set equal to zero.

The reason for the choice of such an observation-function is clear: a most intriguing property of spiral waves is that despite being 
propagating waves affecting all accessible space, they behave as effectively localized particles-like objects \cite{Biktashev2003}. 
The particle-like behavior of spirals corresponds to an effective localization of so called \emph{response functions} 
\cite{Henry2002,Biktashev2010}. The asymptotic theory of the spiral wave drift \cite{keener1988dynamics} 
is based on the idea of summation of elementary responses of the spiral wave core position and rotation phase to elementary 
perturbations of different modalities and at different times and places. This is mathematically expressed in terms of 
the response functions. They decay quickly with distance from the spiral wave core and are almost equal to zero in the region where the spiral 
wave is insensitive to small perturbations.

The numerical results for the sup-optimal control ($\kappa = 0$) and for the sparse sub-optimal control ($\kappa = 1$) are depicted 
in Fig. \ref{F:2b} and Fig. \ref{F:2c}, respectively. One notices that the prescribed spiral tip trajectory is realized for both 
choices for the sparsity parameter $\kappa$, viz., $\kappa = 0$ and $\kappa = 1$. The traces of the spiral tip is indicated by 
the solid lines in both panels. Since the spiral tip rotates rigidly around the spiral core which moves itself on a straight 
line according to $\vec{\phi}(t)$, one observes a periodic motion of the tip in the $x_2$--$t$--plane. In addition, the area 
of non-zero control (colored areas) is obviously much smaller for non-zero sparsity parameter $\kappa$ compared to the case 
$\kappa = 0$ , cf. Fig. \ref{F:2b} and Fig. \ref{F:2c}. However, in this example we observed that the amplitude of the sparse control to be twice as large compared to optimal control ($\kappa=0$).

\FloatBarrier

\subsection{Second-Order Optimality Conditions and Numerical Stability}
\index{Second-Order Optimality Conditions}
To avoid this subsection to become too technical, we only state the main results from Ref. \cite{casas_ryll_troeltzsch2014b}. 
We know for an unconstrained problem with differentiable objective-functional that 
it is sufficient to show $F^\prime(\bar f) = 0$ and $F^{\prime\prime}(\bar f) > 0$ to derive that $\bar f$ is a local minimizer 
of $F$, if $F$ were a real-valued function of one real variable. More details about the importance of second order optimality 
conditions in the context of PDE control can be found in 
Ref. \cite{casas_troeltzsch_ssc}.

In our setting, considering all directions $h \neq 0$ out of a certain so-called critical cone $C_f$, the condition for $\nu > 0$ reads 
\[
 F^{\prime\prime}(\bar f)h^2 > 0.
\]
Then, $\bar f$ is a locally optimal solution of \Pbsp. The detailed structure of $C_f$ is described in Ref. \cite{casas_ryll_troeltzsch2014b};
also the much more complicated case $\nu = 0$ is discussed there.

The second-order sufficient optimality conditions are the basis for interesting questions, e.g. the stability 
of solutions for perturbed desired trajectories and desired states \cite{casas_ryll_troeltzsch2014b}. Moreover, we study the limiting case of Tikhonov parameter $\nu$ tending to zero.

\subsection{Tikhonov parameter tending to zero}

In this section, we investigate the behavior of a sequence of optimal controls and the 
corresponding states as solutions of the problem \Pbsp as $\nu \to 0$. 
For this reason, we denote our control problem \Pbn, the associated optimal control with $\bar f_\nu$, and its associated 
states with $(\bar u_\nu,\bar v_\nu)$ for a fixed $\nu \ge 0$. Since $\Uad$ is bounded in $L^\infty(Q)$, any sequence of 
solutions $\{\bar f_\nu\}_{\nu > 0}$ of \Pbn has subsequences 
converging weakly$^*$ in $L^\infty(Q)$. For a direct numerical approach, this is useless but we can deduce interesting consequences 
of this convergence using second order sufficient optimality conditions.

Assume that the second order sufficient optimality conditions of Ref. \cite[Theorem 4.7]{casas_ryll_troeltzsch2014b} are satisfied.
Then, we derive a H\"older rate of convergence for the states
\begin{equation}
\displaystyle \lim_{\nu \to 0}\frac{1}{\sqrt{\nu}}\left\{\|\bar u_\nu - \bar u_0\|_{L^2(Q)} + \|\bar v_\nu - \bar v_0\|_{L^2(Q)}\right\} = 0,
\label{E4.3}
\end{equation}
with $(\bar u_\nu,\bar v_\nu) = G(\bar f_\nu)$ and $(\bar u_0,\bar v_0) = G(\bar f_0)$. We should mention that this estimate is fairly pessimistic. All of our numerical tests show that the convergence rate is of order $\nu$, i.e., we observe a Lipschitz rather than a H\"older estimate \cite{casas_ryll_troeltzsch2014b}. As mentioned in Ref. \cite{casas_ryll_troeltzsch2014b}, it should also be possible to prove Lipschitz stability and hence to confirm the linear rate of convergence for $\nu \rightarrow 0$ with a remarkable amount of effort.

\subsection{Example 3: Sparse Optimal Control with Tikhonov parameter tending to zero}  \label{SS3.2}

Finally, we consider a traveling pulse solution in the FitzHugh-Nagumo system in one spatial dimension $N=1$. Here, the 
limiting case of vanishing Tikhonov parameter, $\nu = 0$, is of our special interest. We observe that Newton-type methods 
yield very high accuracy even for very small values of $\nu > 0$. This allows us to study the convergence behavior 
of solutions for $\nu$ tending to zero as well.

Following Ref. \cite{casas_ryll_troeltzsch2014b} and in contrast to the last example in Sect.~\ref{SS3.1}, we solve the full 
forward-backward-system of optimality. We stress that this is numerically possible solely for non-vanishing value of $\nu$. However, we constructed examples where
an exact solution of the optimality system for $\nu = 0$ is accessible as shown in Ref. \cite[Section 5.3]{casas_ryll_troeltzsch2014b}. 
In this sequel, our reference-solution, denoted by $\bar u_\text{ref}$, will be the solution of \Pbn for $\nu := \nu_\text{ref} = 10^{-10}$.
For smaller values the numerical errors do not allow to observe a further convergence. The distance 
$\|\bar u_\nu - \bar u_\text{ref}\|_{L^2(Q)}$ stagnates between $\nu = 10^{-10}$ and $\nu_\text{ref} < 10^{-10}$.

\begin{figure}[ht]\centering
  \subfloat[\label{F:3a}]{\includegraphics[width=0.333\textwidth]{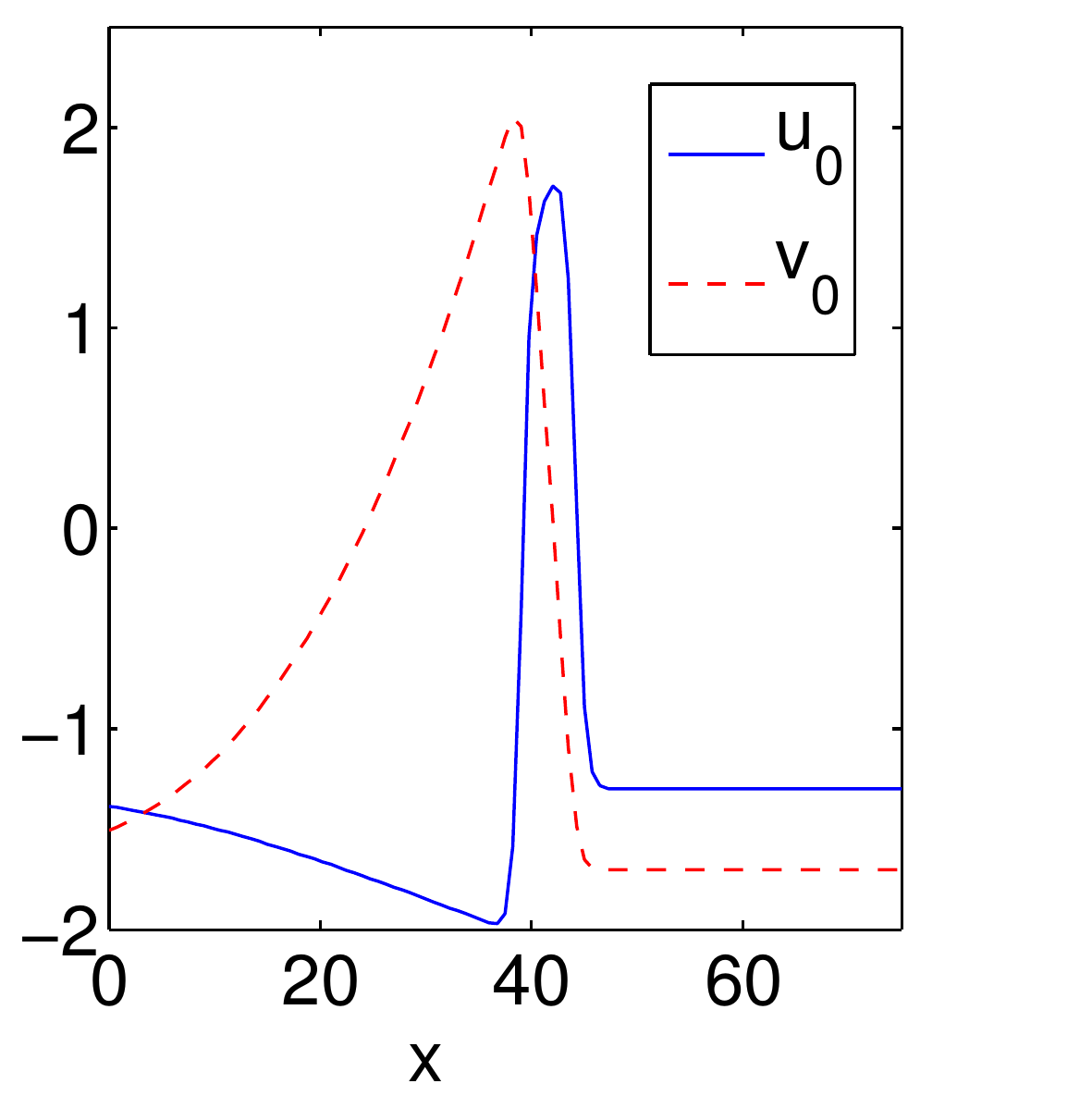}}\hspace*{-7.5pt}
  \subfloat[\label{F:3b}]{\includegraphics[width=0.333\textwidth]{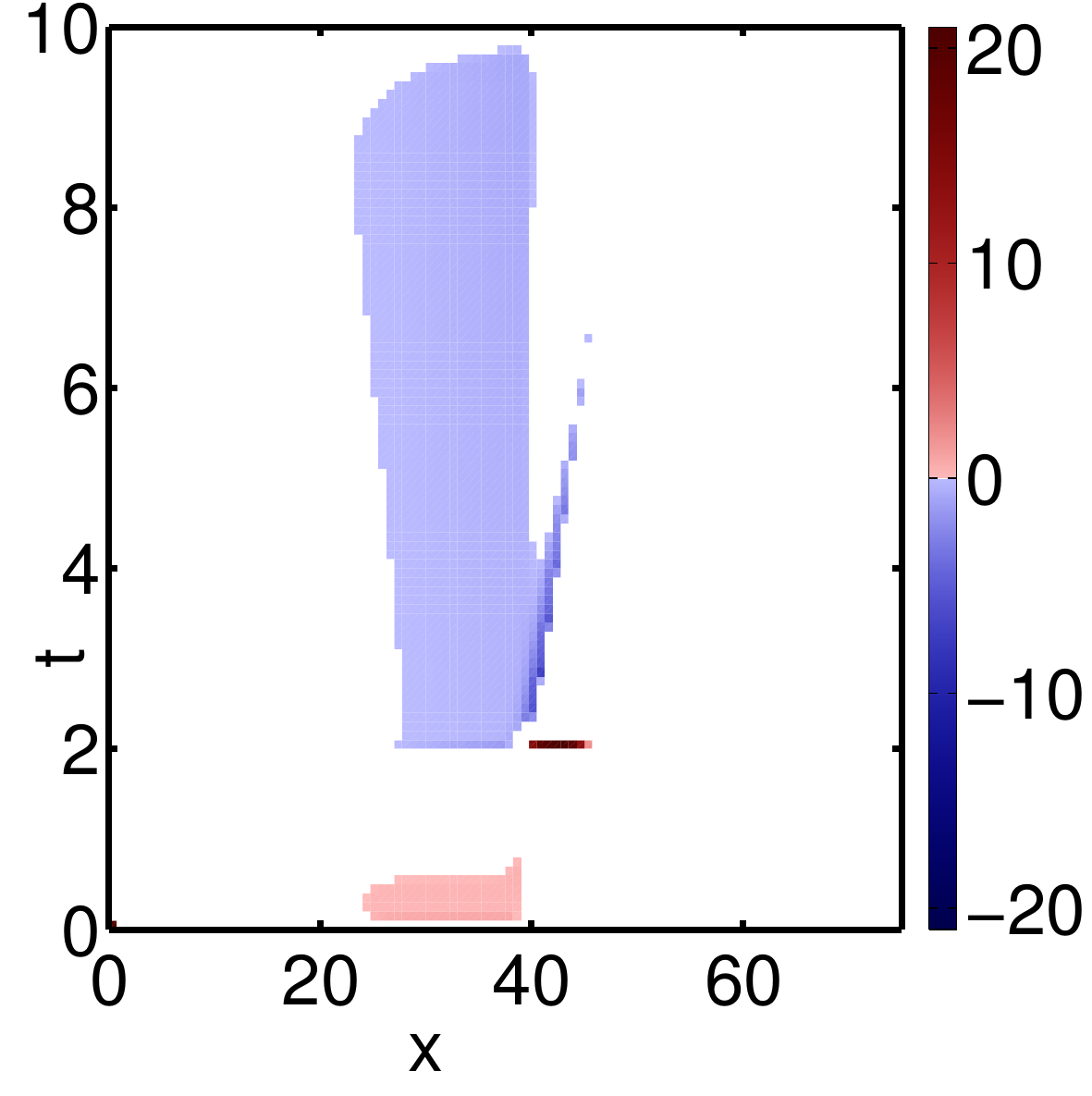}}\hspace*{7,5pt}
  \subfloat[\label{F:3c}]{\includegraphics[width=0.333\textwidth]{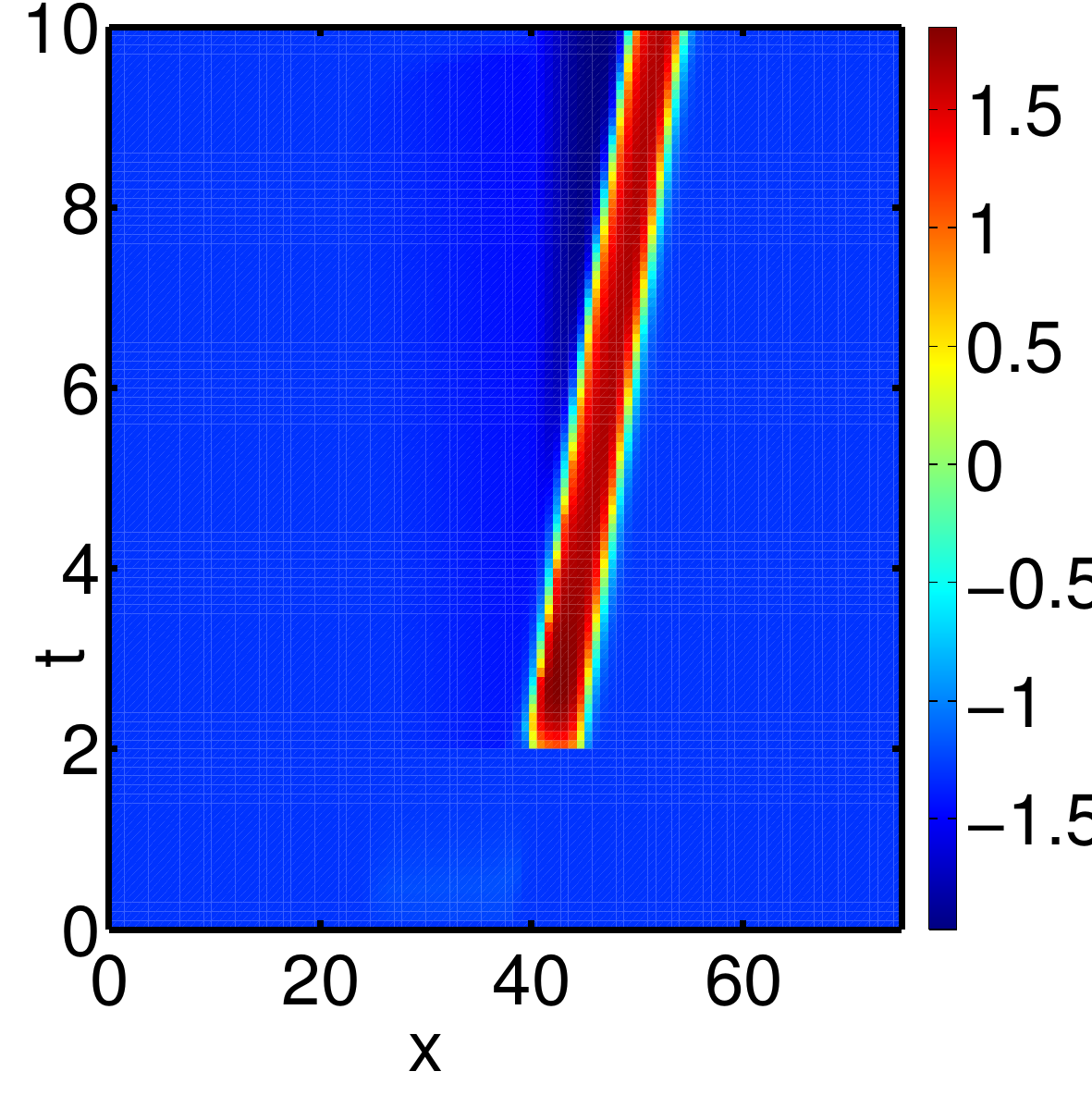}}
  \caption{\textbf{(a)} Segment of a traveling pulse solution $(u_0, v_0)^T$ in the uncontrolled FitzHugh-Nagumo 
  system, \eq{E1.1} with $f = 0$. \textbf{(b)} Numerically obtained sparse optimal control solution $\bar f_\nu$ for almost vanishing 
  Tikhonov parameter, $\nu = 10^{-10}$, and \textbf{(c)} the associated optimal state $\bar u_\nu$. The amplitude of the control signal is color-coded.
  The kinetic parameter values are set to $\alpha = 1$, $\beta = 0$, $\gamma = 0.33$, $\delta = -0.429$, and 
  $R(u) = u(u-\sqrt{3})(u+\sqrt{3})$.}
  \label{F:3}
\end{figure}

Next, we treat the well-studied problem of pulse nucleation \cite{mikhailov1983stochastic,Idris2008} by sparse optimal control. 
We aim to start and to stay in the lower HSS for the first two time-units, i.e., $u_d(\vec{x},t) = -1.3$ for $t \in (0,2)$. 
Then, the activator state shall coincide instantaneously with the traveling pulse solution $u_\text{nat}$, i.e., $u_d(\vec{x},t) = u_\text{nat}(\vec{x},t-2)$. 
To get the activator profile $u_\text{nat}$, we solve \eq{E1.1} for $f = 0$ and its profile is shown in Fig.~\ref{F:3a}.
% The studied optimal control problem is as follows: We aim to start and - for the first two time-units - to stay in the lower homogeneous 
% steady state (HSS), i.e., $u_d(\vec{x},t) = -1.3$ for $t \in (0,2)$. Then, the state shall jump instantaneously into a 
% natural developed traveling pulse solution, denoted by $u_\text{nat}$, i.e. $u_d(\vec{x},t) = u_\text{nat}(\vec{x},t-2)$. 
% To get a solution $u_\text{nat}$, we solve \eq{E1.1} for $f = 0$ and a suitable initial state as shown in Fig.~\ref{F:3a}.

In our optimal control algorithms, we set the parameters to $\Omega = (0,75)$, $T = 10$, $\alpha = 1$, $\beta = 0$, $\gamma = 0.33$, 
and $\delta = -0.429$. Moreover, here we use a slightly different nonlinear reaction kinetics $R(u) = u(u-\sqrt{3})(u+\sqrt{3})$ 
in \eq{E1.1} but this does not change the analytical results. The upper and lower bounds for the control $f$ are set to very 
large values, viz. $f_a = -100$ and $f_b = 100$. In addition, the coefficients in \eq{eq:JFHN} are kept fixed, viz., $c_d^U = 1$ 
and $c_T^U = c_d^V = c_T^V = 0$.

Our numerical results obtained for a sparse optimal control $\bar f_\nu$ acting solely on the activator $u$, 
cf. \eq{E1.1}, are presented in Fig. \ref{F:3b} and Fig. \ref{F:3c}. In order to create a traveling pulse solution 
from the HSS $u_d = -1.3$, the optimal control resembles a step-like excitation with high amplitude at $x \simeq 40$. 
Since the Tikhonov parameter is set to $\nu = 10^{-10}$, large control amplitudes are to be expected and indicate that in 
the unregularized case, even a delta distribution might appear. Because this excitation is supercritical a new pulse will nucleate. 
In order to inhibit the propagation of this nucleated pulse to the left, the control must act at the back of the pulse as well. 
Thus, we observes a negative control amplitude acting in the back of the traveling pulse. We emphasize that the desired shape of 
the pulse is achieved qualitatively. The realization of the exact desired profile can not be expected due to a non-vanishing sparse parameter $\kappa = 0.01$. Even for this respectively small value, the sparsity of the optimal control shows. 

Since the displayed control and state are computed for an almost vanishing value $\nu = 10^{-10}$, 
we take the associated state as reference-state $\bar u_\text{ref}$ in order to study the dependence of the distance 
$\|\bar u_\nu - \bar u_\text{ref}\|_{L^2(Q)}$ on $\nu > 0$. From Table~\ref{T3}, 
one notices the already mentioned Lipschitz-rate of convergence for decreasing values 
$\nu > 0$, $\|\bar u_\nu - \bar u_\text{ref}\|_{L^2(Q)} \propto \nu$. This observation is consistent 
to results from \cite{casas_ryll_troeltzsch2014b} for various other examples.
\begin{table}[ht]\centering\small
  \caption{The comparison of the distance $\|\bar u_\nu - \bar u_\text{ref}\|_{L^2(Q)}$ between the numerically obtained 
  states $\bar u_\nu$ and the numerically obtained reference-solution $\bar u_{\text{ref}}$, computed
  for $\nu = 10^{-10}$, for decreasing values of $\nu > 0$.}
  \begin{tabular}{|c||c|c|c|c|c|c|c|}
  \hline
  $\nu$ & 1E-3 & 1E-4 & 1E-5 & 1E-6 & 1E-7 & 1E-8 & 1E-9\\
  \hline
  $\|\bar u_\nu - \bar u_\text{ref}\|_{L^2(Q)}$ & 1.58E-1 & 1.16E-2 & 1.33E-3 & 1.35E-4 & 1.35E-5 & 1.34E-6 & 1.31E-7 \\
  \hline
  \end{tabular}
  \label{T3}
\end{table}

% \FloatBarrier
\section{Conclusion}\label{S4}

Optimal control of traveling wave patterns in RD systems according to a prescribed desired
distribution is important for many applications. 

Analytical solutions to an unregularized optimal control problem can be obtained with ease 
from the approach presented in Sect. \ref{S1}. In particular, the control signal can be obtained without full knowledge about the underlying
nonlinear reaction kinetics in case of position control. Moreover, they are a good initial guess for the numerical solution of 
regularized optimal control problems with small regularization parameter $\nu > 0$, thereby achieving 
a substantial computational speedup as discussed in Sect. \ref{S2-3}. 
Generally, the analytical expressions may serve as consistency checks for numerical optimal control algorithms.

For the position control of fronts, pulses, and spiral waves, the control signal is spatially localized. By applying sparse optimal 
position control to reaction-diffusion systems, as discussed in Sect. \ref{S3}, the size of the domains with non-vanishing control signals can be further decreased. 
Importantly, the method determines sparse controls without any a priori knowledge about restrictions to certain subdomains. 
Additionally, sparse control allows to study second order optimality conditions that are not only interesting from the theoretical 
perspective but also for numerical Newton-type algorithms.

% Considering problems of position control, both analytic and optimal control provide very satisfying results and verify 
% each other as shown in Sect. \ref{S2-3}. 
% 
% Although there are some strict assumptions made in \ref{S1} on smoothness and fitting of the given data to 
% derive analytic controls, these assumptions seem to be quite natural for many applications. 
% Controls gained this way do not only provide the best results in approximating a desired trajectory 
% in the sense of \eqref{eq:Distdist}, they are also easy to compute and might speed up the numerical 
% calculation of the computational very expensive optimal control significantly - depending on the concrete problem and algorithm. 
% 
% However, the method prescribed in Sect. \ref{S2} is in some sense a more universal tool. 
% In addition to ignoring the assumptions from before, this approach is applicable as well for many restrictions on the control.
% 
% As also shown in Sect. \ref{S2-3}, for problems of position control, the resulting control signal is localized. 
% With the method of sparse optimal control, cf. Sect. \ref{S3}, this effect can be increased. 
% Moreover, this approach allows to investigate second order optimality conditions that are not only of theoretical 
% interest but among others are also fundamental for some numerical Newton-type algorithms.
% 

% \FloatBarrier

\bibliographystyle{spphys_nodoiurl}
\bibliography{litbank}

%\printindex
\end{document}